\documentclass[12pt,preprint]{article}
\pdfoutput=1 
\usepackage{jheppub} 
\usepackage{gensymb}
\usepackage{graphicx,dcolumn,bm,epsfig,cancel,hyperref,float}
\usepackage{multirow}
\usepackage{amsmath}
\usepackage{amssymb, times}
\usepackage[utf8]{inputenc}
\newcommand{\comment}[1]{{}}
\numberwithin{equation}{section}
\def\beq{\begin{align}}
\def\eeq{\end{align}}
\newcommand{\bi}{\begin{itemize}}
\newcommand{\ei}{\end{itemize}}
\newcommand{\ben}{\begin{enumerate}}
\newcommand{\een}{\end{enumerate}}
\newcommand{\be}{\begin{equation}}
\newcommand{\ee}{\end{equation}}
\newcommand{\bea}{\begin{eqnarray}}
\newcommand{\eea}{\end{eqnarray}}

\usepackage{tikz}
\usepackage{subcaption}

\newcommand{\ud}{{\rm d}}

\def\bec{\begin{center}}
\def\eec{\end{center}}
\def\beq{\begin{eqnarray}}
\def\eeq{\end{eqnarray}}

\usepackage{color}
\usepackage{ifthen}
\newboolean{editorial}
\setboolean{editorial}{true}
\newcommand{\editorial}[2]{\ifthenelse{\boolean{editorial}}{\textcolor{red}{[\textsf{\textbf{{#1}}}: }\textcolor{blue}{\textsf{{#2}}}\textcolor{red}{]}}{}}

\usepackage{xcolor}

\author[a, d]{Leia Price,}
\author[b]{Kuver Sinha,}
\author[c]{Robert Wiley Deal}

\affiliation[a]{\footnotesize Leinweber Institude for Theoretical Physics, Department of Physics, University of Michigan, Ann Arbor, MI, 48109, USA}
\affiliation[b]{\footnotesize Department of Physics and Astronomy, University of Oklahoma, Norman, OK, 73019, USA}
\affiliation[c]{\footnotesize Department of Physics, University of Wisconsin, Madison, WI, 53706, USA}
\affiliation[d]{\footnotesize Center for Theoretical Physics---a Leinweber Institute, Department of Physics, Massachusetts Institute of Technology, Cambridge, MA 02139, USA}

\emailAdd{lyprice@mit.edu}
\emailAdd{kuver.sinha@ou.edu}
\emailAdd{wileydeal@wisc.edu}

\title{Global Asymptotics, the Swampland  Conjectures, and Preheating of String Moduli}

\abstract{
While the cosmological implications of the  Swampland Conjectures are usually discussed in the context of inflationary model building,
they also have implications for the violent, non-adiabatic dynamics that can follow inflation or any displacement of string moduli. We study this question through the lens of self-resonant preheating, where we point out that two Swampland motivated structures play central roles. The first is the local curvature of the potential: the tachyonic branch of the refined de Sitter Conjecture singles out the kind of negative curvature that can drive tachyonic amplification. 
We show, however, that local curvature data is not enough; the large field asymptotics of the potential determines how the modulus samples the unstable region. 
Thus, plateaus, barriers, and runaways can lead to different resonance efficiencies; we study tachyonic resonance for bulk moduli in LVS and KKLT compactifications, as well as for typical blow-up moduli potentials and $\alpha$-attractor models. The second Swampland motivated structure pertains to the tower of light states predicted by the Swampland Distance Conjecture. Modeling a finite subset of such states as
an effective stochastic environment within which a string modulus preheats, we find that the light states mainly reshape existing resonance bands by smearing, shifting, and mildly seeding instabilities, rather than opening a robust new reheating channel. Our results suggest that Swampland physics affects preheating  by controlling  both the deterministic curvature structure of the
potential as well as stochastic corrections from emergent light states.
}

\begin{document}

\maketitle

\section{Introduction}

The Swampland program has sharpened the expectation that not every seemingly consistent low energy effective field theory (EFT) admits a UV completion in quantum gravity.  For general reviews, we refer to \cite{Palti:2019pca, vanBeest:2021lhn, Ooguri:2006in, Ooguri:2018wrx}; for applications to various aspects of  cosmology, we refer to \cite{Agrawal:2018own}.

Among the most well developed of the Swampland conjectures are $(1)$ the refined de Sitter Conjecture, which constrains local curvature properties of potentials in string theory and $(2)$ the Swampland Distance Conjecture (SDC), which associates parametrically large field space excursions to the appearance of an exponentially light tower of states. 
These conjectures have been explored extensively in the context of inflationary model building, where they constrain field ranges, the possibility of slow-roll, and the viability of metastable de Sitter-like phases \cite{Agrawal:2018own, Scalisi:2018eaz, Reece:2022soh, Bravo:2020wdr, Lust:2023zql, Garg:2018reu}.

In this work we ask a complementary question: how do Swampland considerations constrain the post-inflationary dynamics of moduli, in particular preheating and related non-perturbative energy transfer? This question is natural for at least two reasons.  First, moduli are ubiquitous in string compactifications and generically undergo coherent oscillations after being displaced from their minima \cite{Kane:2015jia, Allahverdi:2020bys, Batell:2024dsi}.  Second, the same limits in moduli space that underlie parametric EFT control (weak coupling and large volume) are precisely those where the SDC predicts emergent towers and where the Dine-Seiberg  argument \cite{Dine:1985he}  suggests that potentials tend to run away toward zero. Moreover, the possibility of non-perturbative energy transfer typically depends strongly on the local curvature properties of the potential.
These features are therefore potentially important for preheating: efficient resonance is sensitive to the interplay between the local and global shapes of the potential as well as to the presence of light spectator species that can de-phase or catalyze instability bands.

While these questions can be framed within a full inflationary setting, it is useful to separate the question of reheating or preheating from the  inflationary dynamics. In particular, while the Swampland program sets non-trivial constraints on the inflationary potential, such as its curvature and the possibility of controlled trans-Planckian excursions of the inflaton, it is possible to be agnostic about these aspects and focus entirely on the reheating phase. As such, it is useful to consider scalar fields that undergo reheating dynamics but are not necessarily the inflaton, and that is the philosophy adopted in the current work.

Preheating was originally introduced as an efficient reheating mechanism after inflation \cite{Kofman:1994rk, Kofman:1997yn} and has been studied in several string-motivated inflationary scenarios, including axion monodromy, $\alpha$-attractors \cite{Silverstein:2008sg, McAllister:2014mpa, Kallosh:2013lkr, Carrasco:2015pla, Amin:2014eta, Lozanov:2016hid, Leedom:2024qgr}, and, more recently, Large Volume Scenarios (LVS) \cite{Cicoli:2023opf, Leedom:2024qgr}. Its role for moduli reheating and for the cosmological moduli problem  has also been explored \cite{Kofman:2004yc, Greene:2007sa, Shuhmaher:2005mf, Alsarraj:2021yve, Amin:2018kkg, WileyDeal:2025wgh, Giblin:2017wlo}. In previous work \cite{WileyDeal:2025wgh}, the current authors explored preheating of a modulus into additional fields (notably axions) within explicit compactifications.  Here we focus on a complementary and more minimal question: can a modulus $\varphi$ efficiently preheat through its own self-interactions, and how do Swampland considerations affect this possibility?

The common object underlying both questions is the time-dependent frequency of the
modulus fluctuations $\delta\varphi$. Writing
\begin{equation}
\varphi(t,\mathbf{x})=\bar\varphi(t)+\delta\varphi(t,\mathbf{x}) ,
\end{equation}
the Fourier modes of the fluctuation obey, schematically,
\begin{equation}
\delta\ddot\phi_k+
\left[
k^2+m_{\rm eff}^2(t)+\delta m_{\rm tower}^2(t)
\right]\delta\phi_k=0 .
\label{eq:intro_hill_schematic}
\end{equation}
Here \(k\) is the physical wavenumber, \(m_{\rm eff}^2(t)\equiv V''(\bar\varphi(t))\) is the (deterministic, in our treatment) mass squared parameter generated by the oscillating modulus background, and
\(\delta m_{\rm tower}^2(t)\) denotes possible additional time-dependent contributions from
light spectator states. In the absence of the last term, Eq.~\ref{eq:intro_hill_schematic}
is the usual Hill equation governing self-resonance. The sign, size, and time profile of
\(m_{\rm eff}^2(t)\) determines the deterministic resonance bands; additional light states (which we will model stochastically) can
 perturb these bands by shifting or fluctuating the effective frequency.

Our paper has two intertwined but logically distinct themes, corresponding to these two
mass terms in Eq.~\ref{eq:intro_hill_schematic}. The first concerns the deterministic pump \(m_{\rm eff}^2(t)=V''(\bar\phi(t))\).
We show that global asymptotics controls how the condensate samples tachyonic regions of
the potential. A useful \textit{local} diagnostic for the efficacy of tachyonic preheating is
\begin{equation}
\eta_V \equiv M_{\rm pl}^2\left|\frac{V''}{V}\right| ,
\end{equation}
which measures the curvature available for tachyonic growth. This diagnostic (the ``second slow-roll parameter") is constrained by (the tachyonic option of) the  refined de Sitter Conjecture, and has been  applied to explore Swampland constraints on models of inflation.  Crucially, the full contribution from  tachyonic oscillation depends not only on the local data encapsulated in $\eta_V$, but also on the dwell time inside the tachyonic region and by recovery to a sufficiently stiff, up-curving regime that drives the field back through the instability. Two potentials with similar local inflection data can yield parametrically different tachyonic amplification once their asymptotic tails are taken into account. In string compactifications, these asymptotic behaviors are not freely chosen, but dictated by perturbative and non-perturbative effects. The study of the interplay between local $\eta_V$ data and global asymptotics in the deterministic pump \(m_{\rm eff}^2(t)=V''(\bar\varphi(t))\), and whether tachyonic oscillations make an $\mathcal{O}(1)$ contribution to preheating in an expanding universe, is the focus of Section~\ref{setupgeneral}.

The second theme of our paper concerns the correction \(\delta m_{\rm tower}^2(t)\) in Eq.~\ref{eq:intro_hill_schematic}. Motivated by the
SDC, we ask how a finite set of light tower states can modify
the  Hill equation. Rather than treating the tower as a fully resolved multifield system,
we use a stochastic Hill framework to model its coarse grained effect as cycle-to-cycle
fluctuations of the effective frequency. This stochastic reshaping of the deterministic band
structure is the focus of Section~\ref{lvs-kk-swampland}.

We end the Introduction with a comment on the Swampland inputs used in this paper.
There are two logically distinct roles played by Swampland ideas. First, the deterministic
part of the self-resonance problem is controlled by the curvature term \(V''(\bar\varphi(t))\)
in the Hill equation for \(\delta\varphi_k\). This makes the quantity $\eta_V$ a useful diagnostic of tachyonic amplification in  regions  of positive potential with
\(V''<0\). As mentioned before, this diagnostic is closely related to the tachyonic branch of the refined
de Sitter conjecture, but in this paper we use it as a dynamical indicator of
self-resonance rather than as a consistency condition imposed on the EFT. Second, the
Swampland Distance Conjecture motivates the appearance of light towers at large field
distance, which can modify the same Hill equation through additional time-dependent
mass corrections. We do not take a position here on the de Sitter Swampland conjecture
as a global constraint on stabilized vacua. Our analysis concerns non-adiabatic dynamics
of oscillating moduli in phenomenological stabilized potentials and asks how curvature
structure and emergent light species affect the resonance problem.

In Section~\ref{setupgeneral} we develop our results on global asymptotics: we review
self-resonant preheating for a single scalar condensate, introduce semi-analytical diagnostics of
tachyonic oscillations, and then compare representative asymptotic behaviors
in benchmark potentials relevant for moduli. In Section~\ref{lvs-kk-swampland} we
incorporate SDC-motivated light
towers via a stochastic Hill framework, allowing us to model how emergent species smear, shift, and
de-phase resonance bands, as well as how energy deposition into the KK sector can proceed in this
effective description.

\section{Self-Resonant Preheating: Global Asymptotics}\label{setupgeneral}

We briefly recall the dynamical criteria for self-resonant preheating of a single scalar condensate in an expanding universe. 
Although the discussion is kept general here, we will ultimately be interested in a 
canonically normalized modulus $\varphi$ in a typical type IIB flux compactification setting. After inflation (or any displacement event), $\varphi$ undergoes coherent oscillations about a local minimum $\varphi_0$ with Hubble $H\!\ll\!m_\varphi$. We ask whether self-resonant preheating of $\varphi$, that is, the exponential amplification of its own perturbations $\delta\varphi_k$ driven by the background oscillation, can be efficient in the domain of EFT control.

We begin by recapitulating the origin of Eq.~\ref{eq:intro_hill_schematic}, starting with the action
\begin{equation}
    S
    \supset 
    \int
    d^4x
    \,
    \sqrt{-g}
    \left[ 
        \frac{1}{2}
        \partial_\mu 
        \varphi 
        \,
        \partial^\mu 
        \varphi 
        -
        V(\varphi)
    \right]
    .
\end{equation}
The coherently oscillating homogeneous modulus condensate $\bar\varphi$ has small perturbations, $\varphi(t,\textbf{x})=\bar\varphi(t)+\delta\varphi(t,\textbf{x})$, and we wish to study the production of these $\delta\varphi$.
We consider the Fourier components of the fluctuations,
\begin{align}
    \delta\varphi(t,\textbf{x})
    =
    \int 
    \frac{\ud^3\textbf{k}}{(2\pi)^{3/2}} 
    e^{i\textbf{k}\cdot\textbf{x}}
    \,
    \delta\varphi_k(t)
    \ .
\end{align}
Very generally, in situations where single field self-resonant behavior occurs, the linearized fluctuations obey a Hill/Mathieu-type equation of motion,
\begin{equation} \label{generalhill1}
\delta\ddot\varphi_k + 3H\,\delta\dot\varphi_k + \Big[k^2 + m_{\rm eff}^2(t)\Big]\delta\varphi_k = 0\ \ , 
\quad 
m_{\rm eff}^2(t)\equiv V^{''}\big(\bar\varphi(t)\big),
\end{equation}
with dots denoting derivatives with respect to comoving time $t$, primes denoting derivatives with respect to $\varphi$, and $k=|\textbf{k}|$ is the magnitude of physical wavenumbers.

The equation of motion (EOM) of the zero mode $\bar\varphi$, neglecting expansion, is
\begin{align}
    \ddot{\bar\varphi} + \frac{\text{d}V}{\text{d}\bar\varphi} = 0\ .
\end{align}
The oscillating zero mode of the modulus has a maximum value $\Phi$ that we may also refer to as the amplitude of oscillation, although we will see that the minimum value of $\bar\varphi$ is not necessarily $-\Phi$. 
The resulting oscillatory behavior of the zero mode varies greatly depending on the particular potential and its maximum value $\Phi$.
Some examples are worked out in the following parts of this Section. 
These zero-mode solutions, solved numerically, enter into the EOMs of the $k$-modes $\delta\varphi_k$.
Re-incorporating cosmic expansion under an adiabatic assumption looks like a redshifting amplitude, $\dot\Phi(t)<0$.

\subsection{Semi-analytical Treatment of Tachyonic Oscillations}

The growth of perturbations occurs through two main mechanisms, corresponding approximately to the regions of the potential traversed by the background field. The first (tachyonic oscillation) is when the field passes periodically through the tachyonic region of the potential, in the vicinity of an inflection point. The second (parametric resonance) occurs near the stable minimum of the modulus potential. Both effects are captured through a  Floquet analysis of the Hill's equation in Eq.~\ref{generalhill1}. 

According to Floquet's theorem, the solution to Hill's equation of the form $\delta\ddot\varphi_k+\Omega^2_T(t)\delta\varphi_k=0$, where $\Omega^2_T(t)$ has period $T$, is
\begin{align}
    \delta\varphi_k=c_1P_1(t)e^{\mu_k t}+c_2P_2(t) e^{-\mu_k t}\ ,
    \label{eqn:floq-thm}
\end{align}
where $P_{1,2}(t)$ also have periodicity $T$.
The Floquet exponent (also known as the  ``growth rate'') $\gamma\equiv{\rm Re}(\mu)\geq0$ characterizes the solution, and we will use $\gamma$ to gauge the efficiency of the resonance process. We note that it can be difficult to disentangle and compare the contributions of the two mechanisms (tachyonic oscillations and parametric resonance) without a lattice simulation, which ultimately determines whether or not preheating is successful.
The purpose of this paper, rather, is to guide the categorization of moduli potentials based on their how probable their reheating efficiency, and to discuss the applicability of swampland constraints on this classification. 

We calculate Floquet maps, $\gamma$ plotted over parameter space in the limit of $H=0$, following standard methods such as in Section 2.3.1 of Ref.~\cite{Antusch:2017flz} and Section 3.2.3 of Ref.~\cite{Amin:2014eta}.
Floquet maps include effects of both tachyonic and parametric resonance, but it is hard to extract a time-dependent prediction of preheating behavior with them.
We will take steps in this direction by comparing the full Floquet maps with tachyonic-only approximations.
Other time-dependent additions to Floquet analyses include partially accounting for cosmic expansion by mapping $\gamma/H$, explained in Sec.~\ref{phenomodel} and used throughout the rest of this work.
One may also consider trajectories through parameter space as in Refs.~\cite{Amin:2011hj,PhysRevD.110.123511}.

The tachyonic oscillation contribution, especially, is responsive to the asymptotic behavior of the modulus potential.
Broad, efficient particle production from  tachyonic oscillations typically requires repeated intervals where $m_{\rm eff}^2(t)<0$.
\begin{equation}
\ln\mathcal{G}_k \;\simeq\; \int_0^T {\rm d}t\,
{\rm Re}\left(\sqrt{-k^2-V''(\bar\varphi(t))}\right)\ ,
\label{eq:logG_area}
\end{equation}
again neglecting cosmic expansion, though we may roughly account for its effects by finding $(\ln\mathcal{G})/HT$.
We will compare $\gamma T$ to $\ln\mathcal{G}$ to separate tachyonic vs. parametric resonance.

To better understand the qualitative expectations for preheating efficiency within a given potential, we can rewrite the integrated tachyonic energy per cycle in terms of the zero-mode as

\begin{equation}
\ln\mathcal{G}_k 
\;\simeq\; 
\oint_{\Phi}
{\rm d}\bar\varphi\,
\frac{
{\rm Re}
\left(
\sqrt{(-k^2-V''(\bar\varphi))/V(\bar\varphi)}
\right)
}
{
\dot{\bar\varphi}/\sqrt{V(\bar\varphi)}
}
\label{eq:logG_areaRewritten}
\end{equation}

At this point, we can identify part of the integrand with the second slow-roll parameter
of the potential,
\begin{align}
    \eta_V ={M_{\rm pl}^2}\left|\frac{V''}{V}\right|\,
\end{align}
which serves as a convenient link between the ideas of tachyonic resonance (where $V''<0$) and the swampland conjectures that constrain $\eta_V$.
Briefly revisiting the equation of motion for the zero-mode, 
\begin{equation}
    \ddot{\bar\varphi}
    +
    3H
    \dot{\bar\varphi}
    +
    V^\prime(\bar\varphi)
    =
    0
    \,
    ,
\end{equation}
the following equivalent form can be written:
\begin{equation}
    \frac{d}{dt}
    \left(
        \frac{1}{2}
        \dot{\bar\varphi}^2
        +
        V(\bar\varphi)
    \right)
    =
    -
    3
    H
    \dot{\bar\varphi}^2.
    \label{eq:zeroModeAlternateForm}
\end{equation}
Neglecting Hubble friction, Eq.~\ref{eq:zeroModeAlternateForm} implies that 
$
    \frac{1}{2}
    \dot{\bar\varphi}^2
    +
    V(\bar\varphi)
    \simeq
    \rho_0
    \,
    ,
$
where the initial energy density is set by $\rho_0 = V(\Phi)$.
Thus, when Hubble friction can be neglected, the integrated tachyonic energy per cycle, Eq.~\ref{eq:logG_areaRewritten}, can be expressed as 
\begin{equation}
\ln\mathcal{G}_{k \text{ small}} 
\;\simeq\; 
\oint_{\Phi}
{\rm d}\bar\varphi\,
\frac{
\sqrt{\eta_V(\bar\varphi)}
}
{
 \sqrt{ 2 (V(\Phi) / V(\varphi) - 1) }
}
\,
.
\label{eq:logG_areaLowHubble}
\end{equation}
From this equation, the following features are immediately evident: 
\begin{itemize}
    \item local data: when the potential curvature as defined by the slow roll parameter $\eta_V$ is large (in stark contrast to its conventional use in inflationary models), the preheating efficiency is \textit{enhanced}
    \item global shape: when the zero mode traverses a large potential difference, the preheating efficiency is \textit{diminished}.
\end{itemize}
A similar expression holds when the effect of Hubble friction is restored, with the replacement of the initial energy density of the zero mode to its instantaneous energy density, $V(\Phi) \rightarrow \rho(\varphi)$, and the evolution of the energy density is governed by $\dot{\rho} = -3 H \dot{\bar\varphi}^2$.
The preheating efficiency thus depends crucially on the energy dissipated from the zero-mode within each oscillation -- both from the shape of the potential and from cosmic expansion -- with minimal energy dissipation and a slow velocity $\dot{\bar\varphi}$ per oscillation maximizing the efficiency.

The appearance of $\eta_V$ in Eq.~\ref{eq:logG_areaLowHubble} also clarifies the
connection to Swampland motivated curvature criteria. In regions with \(V>0\) and
\(V''<0\), the tachyonic branch of the refined de Sitter conjecture would require
\begin{equation}
-{M_{\rm pl}^2}\frac{V''}{V}\gtrsim c' ,
\qquad c'\sim {\cal O}(1),
\label{eq:refined_dS_tachyonic_branch}
\end{equation}
which is precisely a lower bound on the curvature diagnostic that enters the tachyonic
gain. We will not impose Eq.~\ref{eq:refined_dS_tachyonic_branch} as a global
consistency condition on the stabilized potentials studied below. Rather, we use
$\eta_V$ as a local diagnostic: large negative curvature relative to the potential height is favorable for tachyonic amplification, while regions with $\eta_V\ll 1$ (which favor slow-roll) tend to suppress the tachyonic contribution.

As is evident from the appearance of both local as well as global data in Eq.~\ref{eq:logG_areaLowHubble}, this diagnostic is necessary but not sufficient. The integrated gain also depends on the
trajectory of the zero mode through the potential, the available field range before hitting a
barrier or asymptotic tail, and the competition with Hubble dilution. Thus potentials with
similar local tachyonic curvature can have parametrically different preheating efficiencies
once their global asymptotics are included. This is the sense in which the large-field shape
of moduli potentials remains central to the analysis below. However, purely at the level of local data, one can make the following statement based on Eqs. \ref{eq:logG_areaLowHubble} and \ref{eq:refined_dS_tachyonic_branch}: \textit{the refined de Sitter Conjecture favors local data in string potentials that facilitate enhanced tachyonic resonance.}

In the remainder of this section, we study potentials with two categories of global behaviors: plateaus and  runaways.
These potentials both feature a minimum at $\varphi = \varphi_0$ (where the modulus is stabilized) and a high barrier for $\varphi \ll \varphi_0$.
On the opposing side of the potential, $\varphi \gg \varphi_0$, plateaus have $V'>0$ and $V\to{\rm constant}>0$ whereas runaways have a peak followed by gentle roll to $V\!\to\!0$.
We begin with a heuristic, phenomenological potential in Section~\ref{phenomodel} to understand some key ideas before moving on to true moduli potentials.
In Sections~\ref{blowup} and ~\ref{infl-potentials}, we study a few examples of plateaus before investigating Dine-Seiberg runaway potentials in Section~\ref{dine-seiberg}.
In each case, we will obtain the growth rates from a full numerical Floquet analysis, and obtain an estimate of the contribution from tachyonic oscillations from a semi-analytical treatment.

\subsection{Phenomenological Potential}\label{phenomodel}

In this section, we study a class of inflection point potentials to use as a benchmark for self-resonant preheating, mimicking the shape of moduli potentials.

\begin{figure}[htb!]
\centering
\includegraphics[width=.9\textwidth]{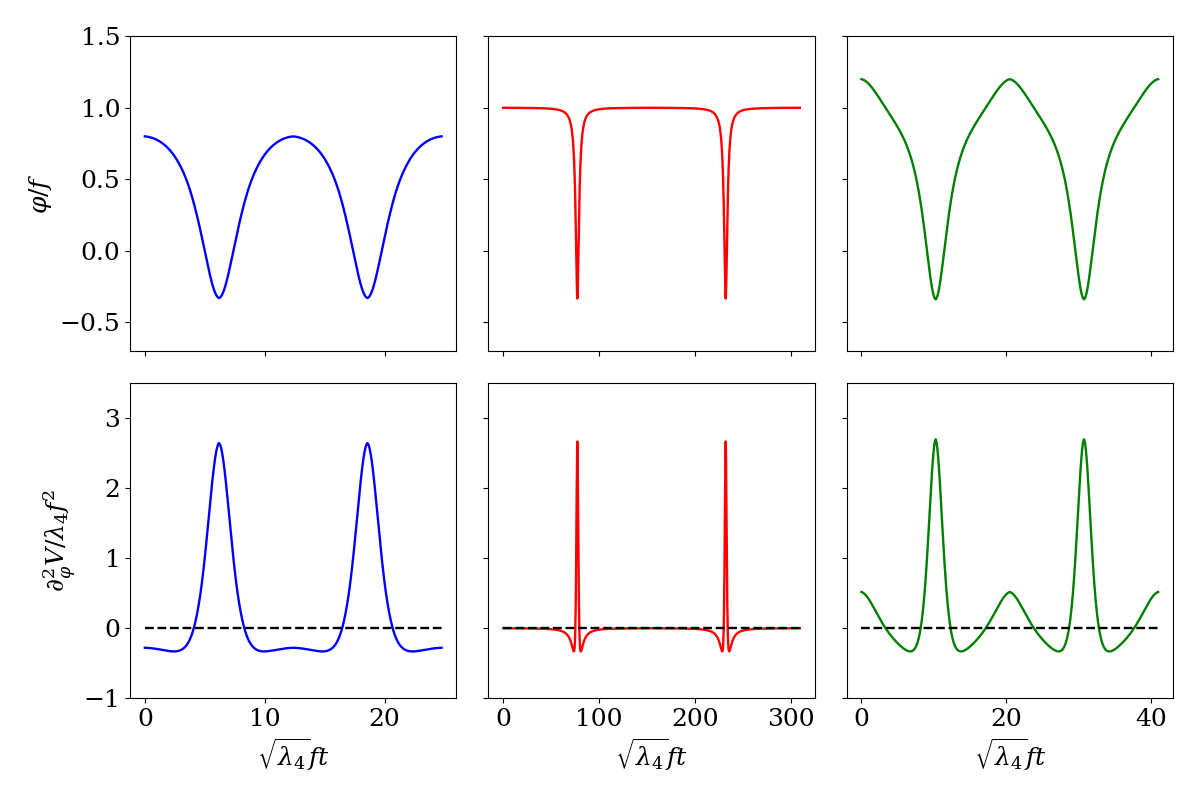}
    \caption{\textit{Top row:} Oscillatory behavior of the  zero mode of the modulus with an amplitude that coincides near the critical point, $\Phi/f=0.8, \, 1.0-10^{-6}, \, \text{and}\, 1.2$ from left to right, with $\Phi$ being the maximum value of the  oscillating zero mode. 
    \textit{Bottom row:} Behavior of $V''$, which is the relevant quantity in the Hill's equation. The time spent in the tachyonic regime, $V'' < 0$, partially determines the success of self-resonant growth.}
    \label{fig:inflection_examples}
\end{figure}

We begin with a general phenomenological parametrization near a critical inflection point (i.e. $V'= V'' = 0$).
We model the neighborhood of the critical point located at $\varphi_\text{crit}\equiv f$ by

\begin{equation}
V(\varphi)\simeq V_0+\frac{\lambda_3}{3}(\varphi-f)^3
+\frac{\lambda_4}{4}(\varphi-f)^4+\cdots,\qquad \lambda_3\neq 0,\;\lambda_4>0\ ,
\label{eqn:pheno-V-shifted}
\end{equation}
and set $\lambda_2=0$ to preserve the desired shape of the potential.\footnote{More specifically, $\lambda_2>\lambda_4f^2/4$ is required to prevent the inclusion of a false vacuum and complications of domain walls and tunneling effects \textit{unless} $\lambda_2=0$. Choosing a $\lambda_2$ above this bound obscures the shape of the potential we are trying to characterize.}
\begin{figure}[htb!]
    \centering
    \includegraphics[width=0.49\linewidth]{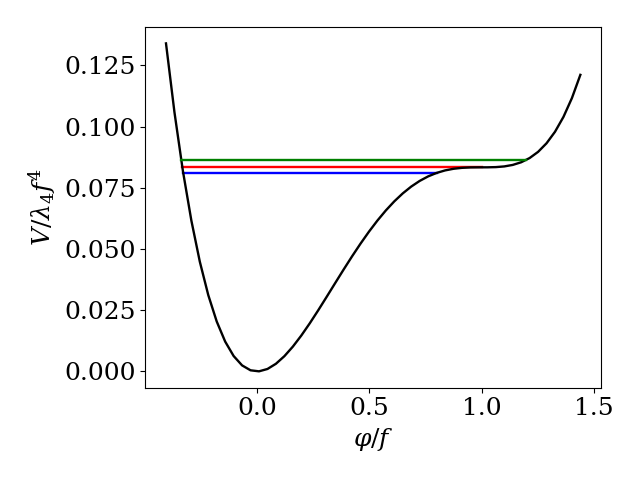}
    \includegraphics[width=0.49\linewidth]{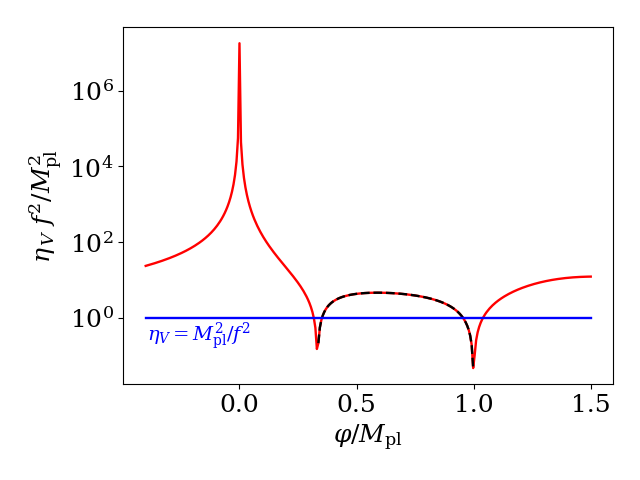}
    \caption{\textit{Left:} Our example of a potential with a critical inflection point at $\varphi/f=1$, corresponding to Eq.~\ref{eqn:pheno-V} with three example oscillation regions $\Phi/f=0.8, \,1, \,\text{and} \,1.2$ in blue, red, and green, respectively.  Here,  $\Phi$ denotes the maximum value of the  oscillating zero mode $\bar\varphi$. 
    The zero mode oscillatory solutions for these regions are shown in Fig.~\ref{fig:inflection_examples}.
    The non-critical inflection point ($V''=0$, $V' \neq 0$) is at $\varphi/f=1/3$, so we expect to see tachyonic resonance for $\Phi/f>1/3$ where $V'' < 0$.
    \textit{Right:} We plot $\eta_V$ normalized by $M^2_{\rm Pl}/f$ , highlighting $V''<0$ in dashed black to emphasize the region where tachyonic self-resonance is possible.
    }
    \label{fig:pheno-potential}
\end{figure}
However, once we constrain the potential further we will find that $m_\varphi^2\neq 0$ (not to be confused with $m_\mathrm{eff}^2(\varphi)=V''(\varphi)$, which is related by $m_\varphi^2=m_\mathrm{eff}^2(0)$); the choices
\begin{align}
    V_0=\frac{\lambda_3}{3}f^3-\frac{\lambda_4}{4}f^4\ \ , \quad f=\frac{\lambda_3}{\lambda_4}
\end{align}
ensure zero vacuum energy at $\varphi_0=0$ for convenience.
The potential then becomes
\begin{align}
    V(\varphi)=\lambda_4 \left(\frac{1}{2}f^2\varphi^2 - \frac{1}{3}2f\varphi^3 + \frac{1}{4}\varphi^4\right)\ .
    \label{eqn:pheno-V}
\end{align}
We see here that $m_\varphi^2=\lambda_4f^2$.
Fig.~\ref{fig:pheno-potential} shows the shape of this potential. 
Fig.~\ref{fig:inflection_examples} demonstrates the oscillatory behavior of the zero mode $\bar\varphi(t)$ for a few choices of $\Phi/f$. 
One can notice the variation in oscillation period $T$ for different $\Phi/f$.

\begin{figure}[htb!]
    \centering
    \includegraphics[width=0.49\linewidth]{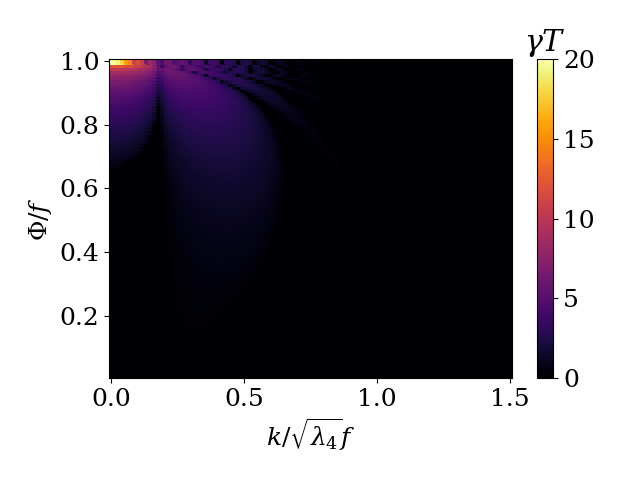}
    \includegraphics[width=0.49\linewidth]{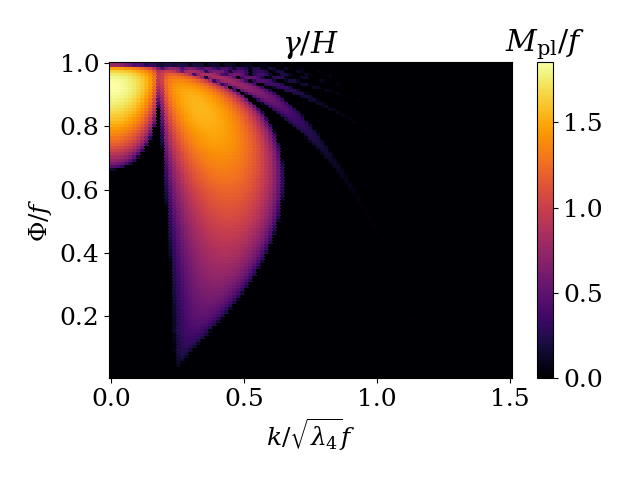}
    \caption{\textit{Left:} Floquet map resulting from the potential displayed in Fig.~\ref{fig:pheno-potential}, showing growth rate per period, $\gamma T$, of Fourier modes $\delta\varphi_k$ being sourced by the zero mode.
    The bright spot highly localized near $\Phi/f=1$ indicates that time spent near the critical point is the most predictive factor of high self-resonance when incorrectly neglecting cosmic expansion.
    \textit{Right:} When comparing the self-resonance rate $\gamma$ to the expansion rate $H$, we uncover a much more interesting structure. 
    Where the potential levels off at $\Phi=f$, resonance is completely inefficient.
    The expected efficient resonance for $\Phi/f>1/3$, the tachyonic region is now evident for $k/\sqrt{\lambda_4}f = k/m_\varphi \gtrsim 0.1$. 
    Interestingly, for $k/m_\varphi\lesssim0.1$, resonance only becomes efficient above $\Phi/f \gtrsim 2/3$.
    }
    \label{fig:quartic-map}
\end{figure}

Our results from the Floquet analysis are presented in  Fig.~\ref{fig:quartic-map}. The Floquet map on the left panel  of Fig.~\ref{fig:quartic-map} shows the growth rates as a function of modulus amplitude $\Phi$ and the wavenumber of the modulus modes \textit{when neglecting expansion}, i.e. $H\to0$.
Although not valid on long time scales, this approximation is sufficiently suitable over the course of an oscillation to provide insight into the underlying physics.
As can be seen from the figure, the growth rate per period is largest for $\Phi/f \simeq 1$, corresponding to oscillations beginning near the critical inflection point (i.e. $V' \simeq V'' \simeq 0$).
The high growth for an amplitude near $\Phi/f=1$ can be explained by considering that $\bar\varphi$ moves slowly through the region where the potential is concave down, $|\varphi/f-\frac{2}{3}|<\frac{1}{3}\,$. 

Of course, the Universe is expanding during the particle production process, and the long dwell time in the tachyonic regime does us no good if $\Phi$ is quickly diluted and thus moved out of a bright spot in the Floquet map.
Rather than include the friction term in Eq.~\ref{generalhill1}, we qualitatively inject cosmic expansion by 1.) recognizing that energy dilution $\rho=V(\Phi)\propto a^{-3}$ decreases $\Phi$ and $k=k_\mathrm{comv}/a$ with time, bringing us in imagined trajectories through the Floquet maps, and 2.) considering $\gamma/H$ rather than $\gamma T$ alone; for efficient resonance we want $\gamma/H\gg1$, or $\gtrsim10$~\cite{Amin:2011hj}.
While not  exact, these two approaches help us to study a wide swath of parameter space, determine the plausibility of the proposed particle production mechanism, and appeal to future work with lattice simulations to provide more concrete estimates of viability.

We show this second modification to the Floquet map on the right panel  of Fig.~\ref{fig:quartic-map}.
It is evident that the long dwell time within the tachyonic regime loses the competition against the expansion timescale for $\Phi/f=1$, and that the vicinity of $\Phi/f\approx0.8-1$ is most hopeful for successful self-resonance in a potential with a critical inflection point.

\begin{figure}[htb!]
    \centering
    \includegraphics[width=0.49\linewidth]{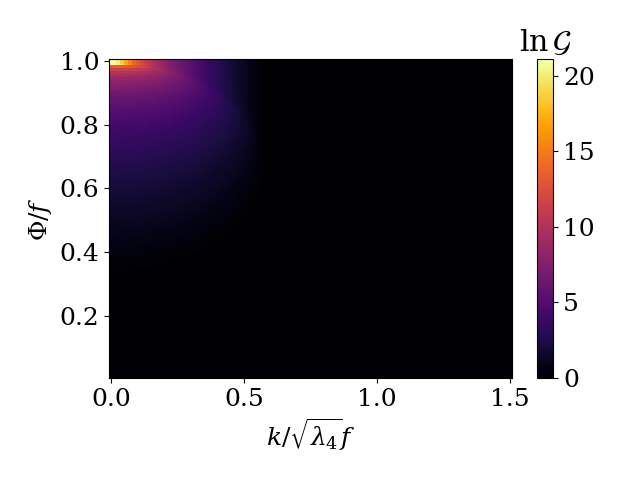}
    \includegraphics[width=0.49\linewidth]{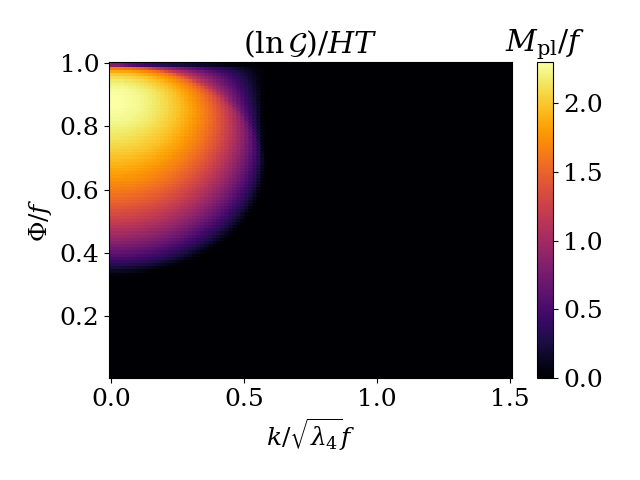}
    \caption{\textit{Left:} Stability map for the potential displayed in Fig.~\ref{fig:pheno-potential},  calculated by integrating the tachyonic effective mass (e.g. using Eq.~\ref{eq:logG_area}), instead of numerically as in Fig. \ref{fig:quartic-map}. Cosmic expansion is ignored.
    \textit{Right:} The tachyonic-integrated growth rate approximation compared to cosmic expansion $H$. Comparing with Fig.~\ref{fig:quartic-map}, we see that this approximation does not account for parametric resonance (any nonzero growth below the inflection point, $\Phi/f<1/3$) and interference effects: the tachyonic approximation slightly overestimates the growth rate values and does not find the band structures present in Fig.~\ref{fig:quartic-map}.
    }
    \label{fig:tach-integrate-map}
\end{figure}

Fig.~\ref{fig:tach-integrate-map} shows the growth rates during the tachyonic oscillation phases only, obtained by the semi-analytical method of integrating over tachyonic regions as in Eq.~\ref{eq:logG_area}, for the case of the potential studied in Eq.~\ref{eqn:pheno-V}. 
Comparing the left panel of Fig.~\ref{fig:tach-integrate-map} to the full Floquet analysis in the left panel of Fig.~\ref{fig:quartic-map}, one can see some agreement, both qualitatively and quantitatively. 
When comparing the right panel of Fig.~\ref{fig:tach-integrate-map} to that of Fig.~\ref{fig:quartic-map}, one can  see that the main differences are (a) the non-zero growth rate in the regime $\Phi/f \sim 0$ due to parametric resonance is apparent in the full Floquet analysis of Fig.~\ref{fig:quartic-map} but is absent in the semi-analytic tachyonic resonance analysis of Fig.~\ref{fig:tach-integrate-map}; and (b) the loss of  narrow band structures that are present in the full Floquet analysis.

We have explored this phenomenological inflection point potential to demonstrate some key ideas of self-resonant particle production.
In the examples comprising the remainder of this paper, the potentials will have tachyonic regions, and we will argue that the large-field asymptotics influence how strongly tachyonic the resonance can be.
In Section~\ref{blowup}, the potential plateaus in the large field limit, while in Section~\ref{dine-seiberg} we encounter the Dine-Seiberg problem, where the potential runs away at large field values.

\subsection{Blow-up Moduli} \label{blowup}

A key goal of this paper is to understand preheating in scalar potentials that arise in controlled corners of string compactifications. Within type IIB flux compactifications, one encounters a large variety of effective single field potentials after stabilization  and restricting to particular trajectories in the modulus space. For relevant reviews, we refer to \cite{Cicoli:2023opf, Baumann:2014nda, McAllister:2007bg, McAllister:2023vgy} and especially \cite{Martin:2013tda}, where a  systematic catalog of such potentials and their viability as inflatons has been compiled. While much of the connection to cosmology has been directed towards inflationary phenomenology, we are interested in the post-inflationary epoch, especially whether these string inspired potentials can support efficient self-resonant preheating.

Among the most ubiquitous and tractable moduli in type IIB are the ``blow-up'' K\"ahler moduli that resolve local singularities and are stabilized by non-perturbative effects. Their potentials arise from an interplay between a non-perturbative superpotential contribution $W\supset A e^{-aT_s}$ and the K\"ahler potential within  Large Volume Scenarios (LVS) \cite{Conlon:2005jm}. After integrating out heavier directions (and, in LVS, holding the overall volume approximately fixed), one obtains an effective potential for the blow-up modulus that is generally exponential in the underlying geometric modulus and non-polynomial in the canonically normalized field. We refer to \cite{Bansal:2026rif} for a recent review and further details.

A representative class of blow-up potentials can be parametrized  by 
\begin{equation}
V(\phi)\;\sim\;V_0\Big[1 - \alpha \, \phi^{p}\,e^{-\,\beta \phi} + \cdots\Big]\,\,,
\label{eq:blowup_generic_form_repeat}
\end{equation}
where $\phi$ is canonically normalized along the chosen trajectory and the ellipsis denotes additional subleading terms, including uplift and loop-induced  contributions, whose precise structure depends on the compactification.
To illustrate the main features, we can take the case $p = \beta = 1$ in the parametrization of Eq.~\ref{eq:blowup_generic_form_repeat}. We also take $V_0 = M^4$, and denote the canonically normalized modulus by $\varphi$. The potential then becomes
\begin{equation}
V(\varphi)=M^4\Bigl[1-\alpha\,\frac{\varphi}{M_{\rm pl}}\,e^{-\varphi/M_{\rm pl}}\Bigr],
\label{eq:KMII_potential}
\end{equation}
where $\alpha$ is a dimensionless parameter.
The inflection point occurs at $\varphi_{\rm inf} \, = \, 2  M_{\rm pl}$,
which may be checked by evaluating
\begin{equation}
V''(\varphi)=\frac{M^4\alpha}{M_{\rm pl}^2}e^{-\varphi/M_{\rm pl}}
\Bigl(2-\frac{\varphi}{M_{\rm pl}}\Bigr),
\end{equation}
so that $V''>0$ for $\varphi<2M_{\rm pl}$ and $V''<0$ for $\varphi>2M_{\rm pl}$ (the tachyonic regime). The potential is thus convex around the minimum and becomes concave only for larger $\varphi$. Moreover, for $\varphi\to\infty$ one has
\begin{equation}
V(\varphi)\longrightarrow M^4,
\end{equation}
so that one obtains asymptotic regions that are a plateau at finite height. 

\begin{figure}[htb!]
    \centering
    \includegraphics[width=0.49\linewidth]{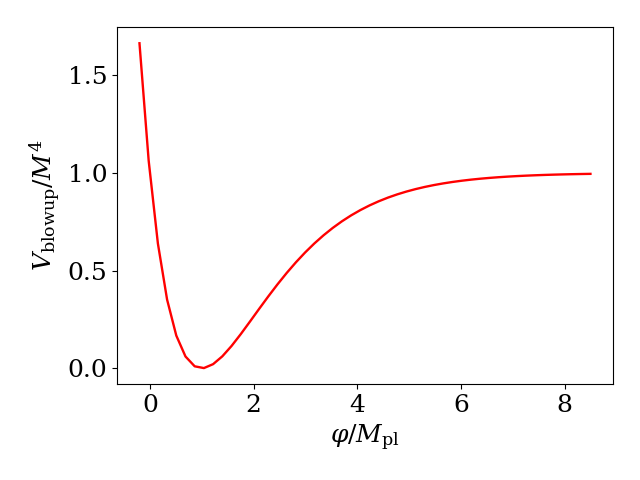}
    \includegraphics[width=0.49\linewidth]{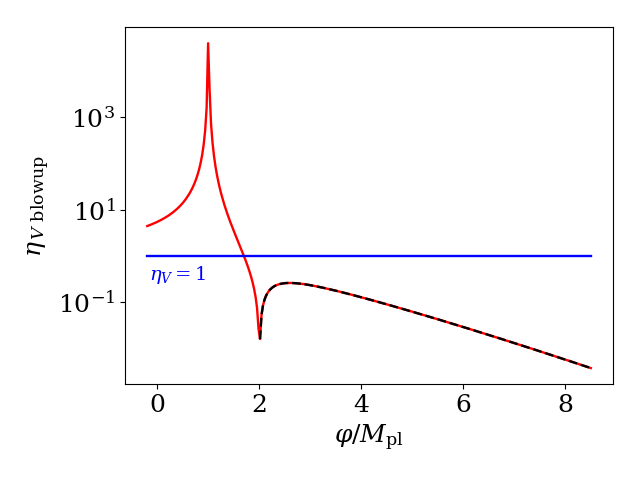}
    \caption{\textit{Left:} The blow-up modulus potential of Eq.~\ref{eq:KMII_potential} with $\alpha=e$. This potential has a minimum at $\varphi_0=M_{\rm pl}$, and plateaus to $V\to M^4$ as $\varphi\to\infty$.
    \textit{Right:} The second slow-roll parameter of the blowup potential, with $V''<0$ highlighted in dashed black. In this region, $\eta_V<1$ in the entirety of this regime predicts that self-resonance of the blowup modulus will be inefficient, and indeed the corresponding growth rates in Fig.~\ref{fig:blowup-result} are small, $\gamma/H\lesssim1$.}
    \label{fig:blowup-V}
\end{figure}

\begin{figure}[htb!]
    \centering
    \includegraphics[width=0.49\linewidth]{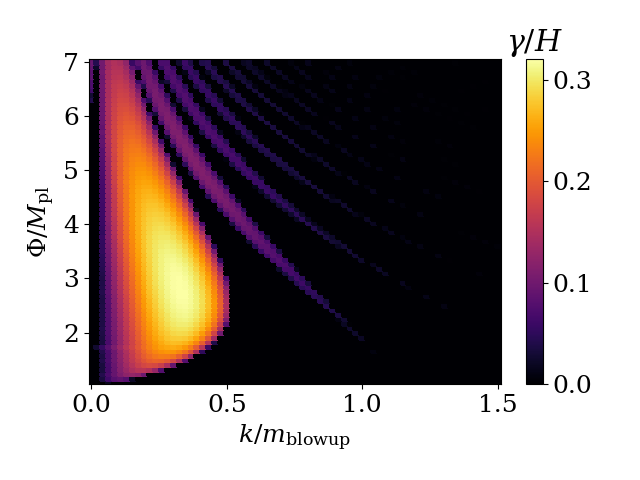}
    \includegraphics[width=0.49\linewidth]{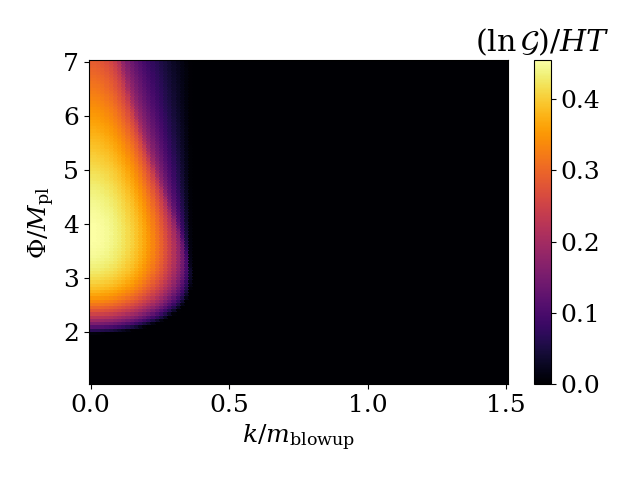}
    \caption{\textit{Left:} Floquet chart for self-resonance within the blow-up modulus potential displayed in Fig.~\ref{fig:blowup-V} as a function of initial modulus amplitude, $\Phi/M_\text{pl}$, and wavenumber $k/m_\mathrm{blowup}$, where $m_\mathrm{blowup}\equiv V_\mathrm{blowup}''(\varphi_0)$.
    \textit{Right:} Growth rates approximated by the integrating the tachyonic mass, as in Eq.~\ref{eq:logG_area}. This allows us to roughly separate the effects of tachyonic resonance from parametric resonance (in this case, $\Phi/M_\mathrm{pl}<2$) and numerically-unearthed interference effects (band structures and damping not present for tachyonic-only effects).}
    \label{fig:blowup-result}
\end{figure}

The potential is plotted in Fig.~\ref{fig:blowup-V}. 
A Floquet analysis for blow-up modes has recently been performed by the authors of \cite{Khan:2021ght} in the context of inflation 
and \cite{Antusch:2017flz} in the context of oscillon formation\footnote{The literature on oscillons is vast; for representative papers, we refer to Refs.~\cite{Allahverdi:2020bys, Lozanov:2016hid, Lozanov:2017hjm}.}.
We repeat and extend this Floquet analysis in Fig.~\ref{fig:blowup-result}, plotting $\gamma/H$ as a function of the modulus amplitude $\Phi$ and wavenumber $k$.


For the semi-analytical results including only the contribution from tachyonic oscillations, shown in the right panel of Fig.~\ref{fig:blowup-result}, it is useful to note that the derivatives of the potential are, schematically,
\begin{equation}
V'(\varphi)\sim V_0\,\mathcal{A}\,e^{-\beta\varphi}\times(\beta\varphi^{p}+\cdots),
\qquad
V''(\varphi)\sim -V_0\,\mathcal{A}\,e^{-\beta\varphi}\times(\beta^2\varphi^{p}+\cdots),
\end{equation}
so the same exponential factor that shapes the potential also controls the magnitude and time variation of $V''(\varphi(t))$ during an oscillation. Whether $V''$ can become negative within the accessible oscillation amplitude depends on the detailed competition among the terms in the potential, which arise from contributions from uplifting, loops, and the non-perturbative superpotential.
This class of  potentials, for general choices of $\beta$ and $p$, therefore  scans over the effective area under the instability in Eq.~\ref{eq:logG_area}. For the particular choice of $p=\beta=1$, we obtain the results shown on the right panel of Fig.~\ref{fig:blowup-result}. We see that there is broad agreement with the full Floquet analysis on the left panel, which also captures the parametric resonance region near $\Phi \sim 0$.

Interestingly, here we find only a small amplification with a growth rate around $\gamma/H \sim 0.3$ or so. 
This result is in agreement with Ref.~\cite{Khan:2021ght}, but seemingly contrasts with the results of Ref.~\cite{Antusch:2017flz}.
The difference in results can be understood via our analysis of the tachyonic gain in Eq.~\ref{eq:logG_areaLowHubble}.
For any plateau model, we expect low $\eta_V$ at sufficiently large values of $\varphi$. 
However, the location of the inflection point of the potential relative to where the plateau region begins may still produce a window containing a sizable $\eta_V$.
Additionally, the factor $V(\varphi_0) / V(\varphi)$ can have significant effect if the ratio is close to 1, where the field velocity is relatively suppressed so that the time in the tachyonic window is enhanced. 
In Ref.~\cite{Antusch:2017flz}, although a different precise form of the potential is used for the blow-up modes, the schematic form produces $V\sim V_0 (1 - \varphi^{4/3} \exp(-a \varphi^{4/3}))^2$ where $a \sim \mathcal{O}(100)$. 
Thus, although the scaling of the potential $V_0$ is irrelevant to our tachyonic gain, the curvature as given by $\eta_V$ becomes \textit{far} larger in the transition region between the plateau and potential minimum than in the case we study, where we have assumed an $\mathcal{O}(1)$ factor within the exponential in our potential.

Blow-up potentials show that  within a broad class of stringy potentials that can support inflationary plateaus or inflection points, although the existence of a tachyonic regime resulting in (some) self-resonant preheating is expected to be generic,  
the plateau can suppress $\eta_V$ and limit the preheating efficiency.
This can be contrasted with the phenomenological model in Section~\ref{phenomodel}, where the plateau $V'=0$ is reached within a finite field displacement because of a higher $\eta_V$.

\subsection{Alpha-attractor Models}  \label{infl-potentials}

A third benchmark class we study is that of $\alpha$-attractors.
As a matter of dynamics, $\alpha$-attractor potentials are known to support particularly efficient self-resonant preheating in a broad region of parameter space \cite{Lozanov:2017hjm, Adshead:2023nhk, Scalisi:2018eaz}. Their defining hyperbolic field space geometry closely resembles the kinetic structure near points at infinite distance in moduli space, precisely the regime emphasized by the Swampland conjectures.

In the simplest $\mathcal{N}=1$ supergravity realization, $\alpha$-attractors arise from a pole in the kinetic term (or equivalently a constant negative curvature in field space). The core structural input of $\alpha$-attractor models is a hyperbolic  kinetic metric,
\begin{equation}
\mathcal{L}_{\rm kin} \;=\; \frac{3\alpha}{4}\,\frac{(\partial\phi)^2}{\phi^2}\,,
\qquad
\phi \;\to\; 0 \;\;\Rightarrow\;\; \varphi \equiv -\sqrt{\tfrac{3\alpha}{2}}\ln\phi \;\; \text{(canonical)}\,,
\end{equation}
which yields an exponential approach to a de Sitter-like plateau in the canonical field $\varphi$ provided the potential is regular in the non-canonical coordinate.

It is interesting to investigate the effect of global asymptotics on self-resonant preheating in these models, especially the contributions from tachyonic oscillations. We refer to \cite{Lozanov:2017hjm, Adshead:2023nhk, Scalisi:2018eaz} for the full calculation of growth rates, and confine ourselves to a semi-analytical discussion. The potentials in alpha-attractor models are of the type  $V\propto |\varphi|^{2n}$ with $n>1$ near the minimum and a plateau beyond a transition point:
\begin{align}
&\text{T-model:}\quad V(\varphi)\propto \tanh^{2n}\!\left(\frac{\varphi}{\sqrt{6\alpha}}\right)
\;\;\Rightarrow\;\; 
V(\varphi)\sim \varphi^{2n}\ \text{near }0,\ \text{plateau for }|\varphi|\gg\sqrt{\alpha}\,;\\
&\text{E-model:}\quad 
V(\varphi)\propto \left(1-e^{-\sqrt{\tfrac{2}{3\alpha}}\varphi}\right)^{2n}
\;\;\Rightarrow\;\;
V(\varphi)\sim \varphi^{2n}\ \text{near }0,\ \text{plateau for }\varphi\gg\sqrt{\alpha}\,.
\end{align}
For $n>1$ (and even for $n=1$ at sufficiently large post-inflationary amplitudes probing the transition region), the condensate undergoes strong self-resonance and fragmentation.

\begin{figure}[htb!]
    \centering
    \includegraphics[width=0.49\linewidth]{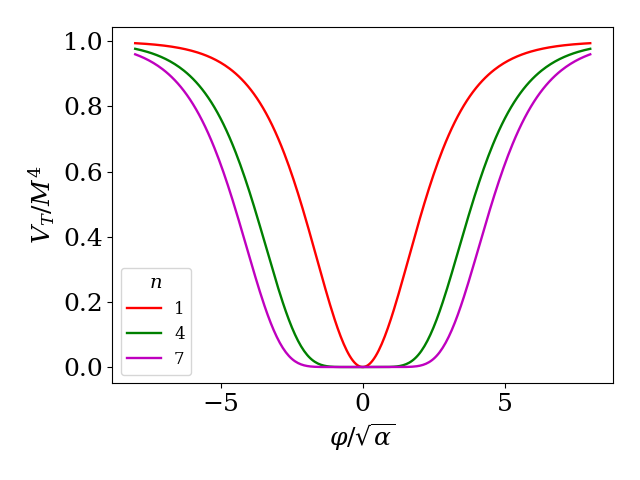}
    \includegraphics[width=0.49\linewidth]{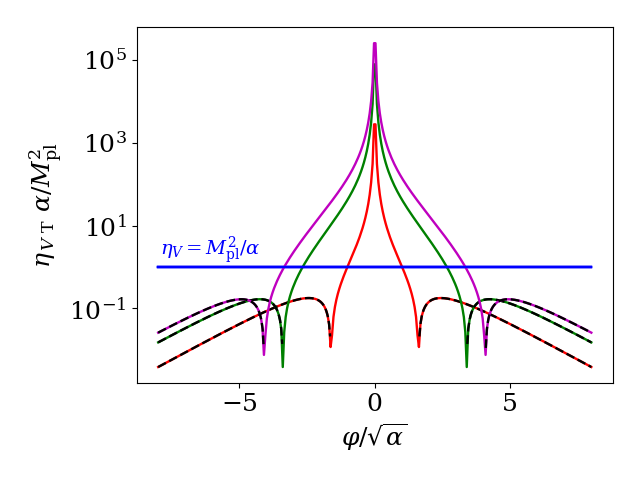}
    \caption{\textit{Left:} The T-model potential for a few choices of $n$. This potential is a two-sided plateau.
     \textit{Right:} Normalized $\eta_V$ of the T-model potential, with $V''<0$ highlighted in dashed black. In this region, $\eta_V<1$ in most of this regime predicts that self-resonance might be efficient depending on $\sqrt{\alpha}/M_{\rm pl}$; the lower this ratio, the more efficient preheating. We refer to Ref.~\cite{Amin:2011hj,Lozanov:2017hjm} for a more thorough investigation of T-models, where oscillon formation is found, indeed for $\sqrt{\alpha}/M_{\rm pl}\ll1$.}
    \label{fig:t-model-potn}
\end{figure}
\begin{figure}[htb!]
    \centering
    \includegraphics[width=0.49\linewidth]{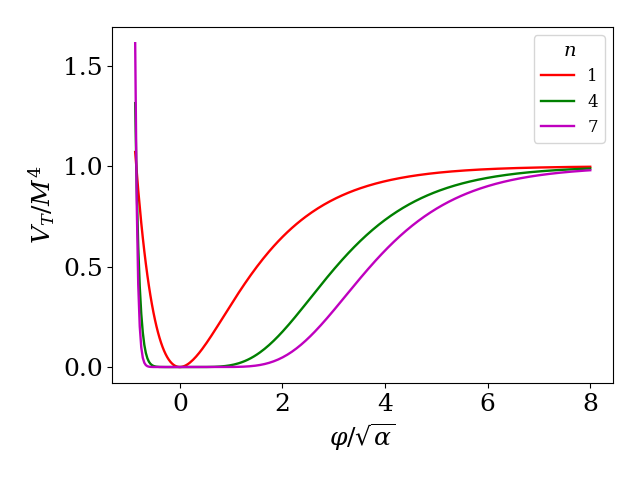}
    \includegraphics[width=0.49\linewidth]{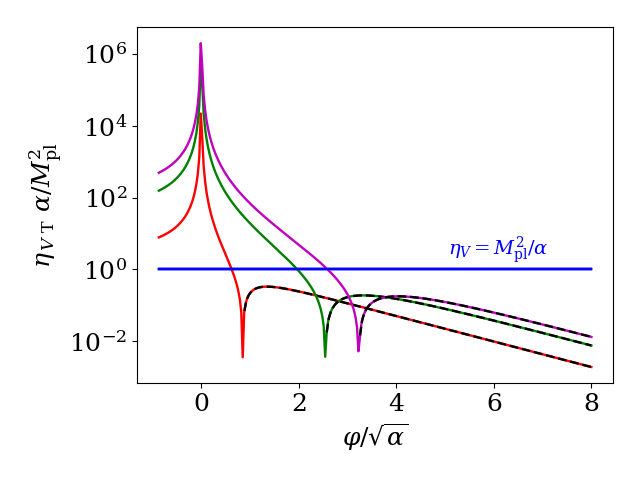}
    \caption{\textit{Left:} The E-model potential for a few choices of $n$.
     \textit{Right:} The second slow-roll parameter of the E-model potential, with $V''<0$ highlighted in dashed black. In this region, $\eta_V<1$ in most of this regime predicts that self-resonance might be efficient depeding on $\sqrt{\alpha}/M_{\rm pl}$. We refer to Ref.~\cite{Amin:2011hj, Lozanov:2017hjm} for a more thorough investigation of E-models, where oscillon formation is found, indeed for $\sqrt{\alpha}/M_{\rm pl}\ll1$.}
    \label{fig:e-model-potn}
\end{figure}

The  most important  parameter for contributions from tachyonic oscillations is the transition scale
\begin{equation}
\varphi_{\rm tr}\;\sim\;\sqrt{\alpha}\,,
\end{equation}
which sets the location where the potential crosses over from polynomial behavior near the minimum to plateau-like behavior at larger field values. We discuss our results for the T-model and E-model in Figs.~\ref{fig:t-model-potn} and \ref{fig:e-model-potn}, respectively.

\subsection{Dine-Seiberg Runaway Potentials}  \label{dine-seiberg}

We finally turn to our last benchmark class: models with asymptotic runaways.

\begin{figure}[htb!]
    \centering
    \includegraphics[width=0.49\linewidth]{potn.png}
    \includegraphics[width=0.49\linewidth]{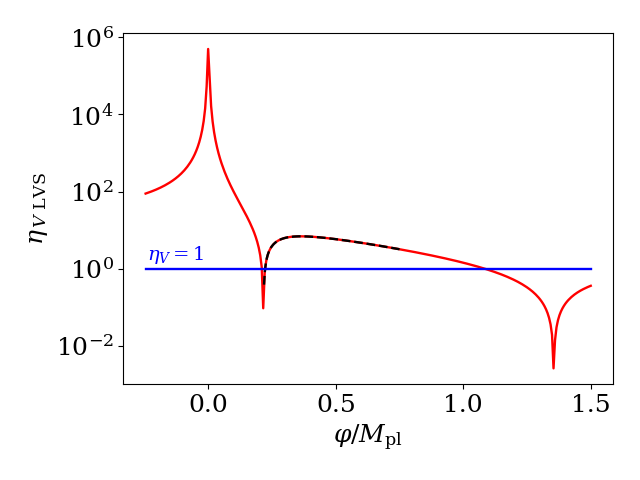}
    \caption{\textit{Left:} The LVS potential, detailed in Appendix~\ref{lvsapp}, with parameters $W_0=A_s=a_s/2\pi=\lambda=\gamma=\xi=1$.
    The coincidental reuse of $\gamma,\lambda,\xi$ appears only in this context; their meanings throughout the rest of the paper are not related to the parameterization of this potential. 
    This potential has a minimum at $\varphi_0=0$, a barrier, and a runaway $V\to0$ as $\varphi\to\infty$.
     \textit{Right:} The $\eta_V$ parameter of the LVS potential, with $\{V''<0,\varphi<\text{peak}\}$ highlighted in dashed black. In this region, mostly $\eta_V>1$ predicts that self-resonance of the LVS modulus will be somewhat efficient but likely incomplete, with the corresponding growth rates in Fig.~\ref{fig:LVS-result} around $\gamma/H\sim1$.
    }
    \label{fig:lvs-potn}
\end{figure}

\begin{figure}[htb!]
    \centering
    \includegraphics[width=0.49\linewidth]{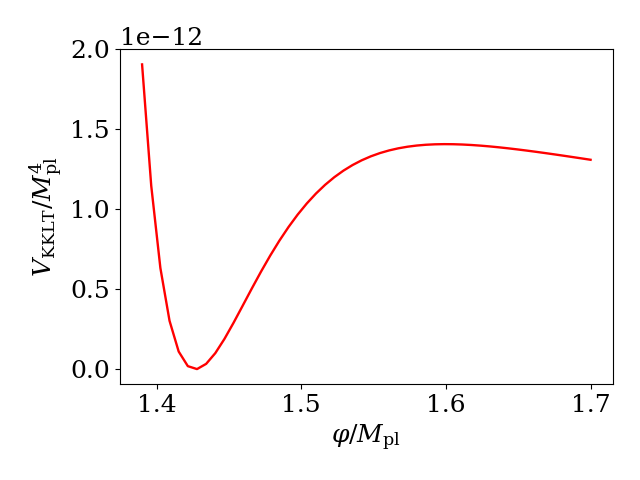}
    \includegraphics[width=0.49\textwidth]{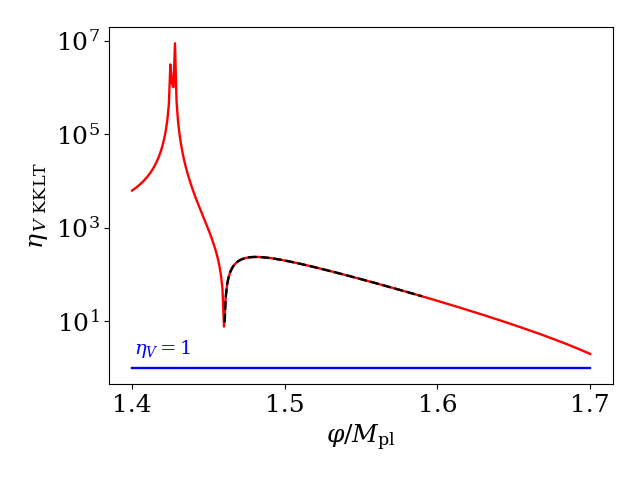}
    \caption{\textit{Left: }The KKLT potential with $W_0 = 10^{-5}, A = 10$ and $a = 2\pi$. For details and definition of parameters, we refer to Appendix~\ref{KKLTblowup}.
    \textit{Right:} The second slow-roll parameter of the KKLT potential, with $\{V''<0,\varphi<\text{peak}\}$ highlighted in dashed black. In this region, $\eta_V\gg1$ predicts that self-resonance of the KKLT modulus will be efficient and worth a numerical investigation for oscillon formation \cite{Antusch:2017flz} and fragmentation completion, with the corresponding growth rates in Fig.~\ref{fig:kklt-result} around $\gamma/H\approx10$.}
    \label{fig:kklt-potn}
\end{figure}

The Dine-Seiberg problem \cite{Dine:1985he}  asserts that in weakly coupled, large radius limits of string theory the scalar potential of a bulk modulus typically runs away toward zero
\begin{equation}
\text{(weak coupling / large radius)} 
\qquad \Longrightarrow \qquad V(\varphi)\;\longrightarrow\; 0^{+}, 
\end{equation}
with $\varphi$ a canonically normalized modulus along an infinite distance trajectory in field space\footnote{For general reviews of flux compactifications, we refer to \cite{Grana:2005jc, Douglas:2006es}.}. In explicit constructions, stabilized vacua at finite volume  appear only if a barrier, typically from non-perturbative effects, fluxes, and uplifting terms, interrupts the generic runaway. We will mostly be working within the context of Large Volume Scenarios (LVS) \cite{Conlon:2005ki, Cicoli:2008va} and KKLT \cite{Kachru:2003aw}, for which these effects have been calculated in detail. We relegate a calculation of the full potential in such scenarios to Appendix \ref{lvsapp} and \ref{KKLTblowup}, where we recast the known form of the bulk volume modulus potential in a form that is useful for our purposes.  The potential for the volume modulus of LVS is depicted in Fig.~\ref{fig:lvs-potn}, while that of KKLT is displayed in Fig.~\ref{fig:kklt-potn}.

\begin{figure}[htb!]
    \centering
    \includegraphics[width=0.49\linewidth]{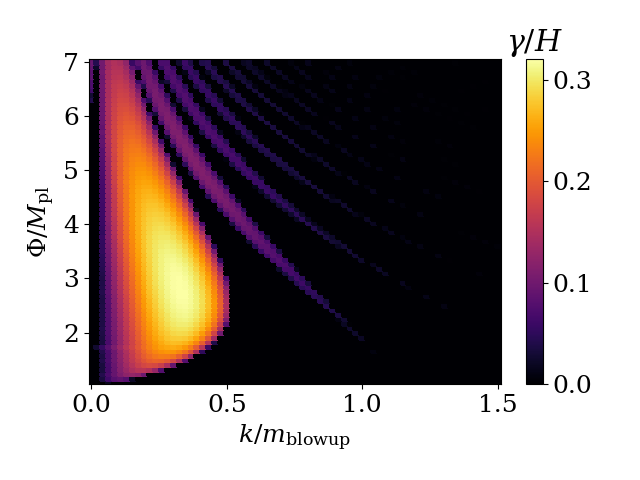}
    \includegraphics[width=0.49\linewidth]{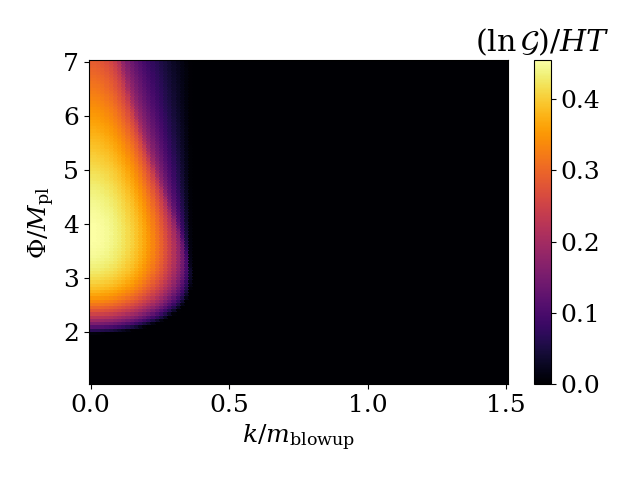}
    \caption{\textit{Left:} Floquet chart for self-resonance the LVS potential displayed in Fig.~\ref{fig:lvs-potn} as a function of modulus amplitude $\Phi/M_\text{pl}$ and wavenumber $k/m_\mathrm{LVS}$, where $m_\mathrm{LVS}^2=V_\mathrm{LVS}''(0)$.
    \textit{Right:} Growth rate approximated by integrating the tachyonic region, as in Eq.~\ref{eq:logG_area}.}
    \label{fig:LVS-result}
\end{figure}

The possibility of preheating the volume modulus into other fields, especially axions, has been addressed by the current authors \cite{WileyDeal:2025wgh} as well as other groups \cite{Leedom:2024qgr}. Our interest in the current work is in self-preheating. An important consequence of the Dine-Seiberg runaway is that in controlled corners of moduli space such as that depicted in Fig.~\ref{fig:lvs-potn}, the potential is generally asymmetric around the minimum. 
One side faces a barrier (uplifted AdS $\to$ dS, or an AdS potential well before uplift); the other side leans toward a runaway, where $V\to 0$ at large field distance. 
This asymptote results in a higher $\eta_V$ in the oscillation region compared to a plateau, which becomes more slow-roll like for larger $\varphi$.
Despite the fact that the modulus must stay on the left side of the peak barrier for the preheating discussion to be relevant, this heightened $\eta_V$ predicts higher preheating efficiency than the plateau potentials.

\begin{figure}[htb!]
    \centering
    \includegraphics[width=0.49\linewidth]{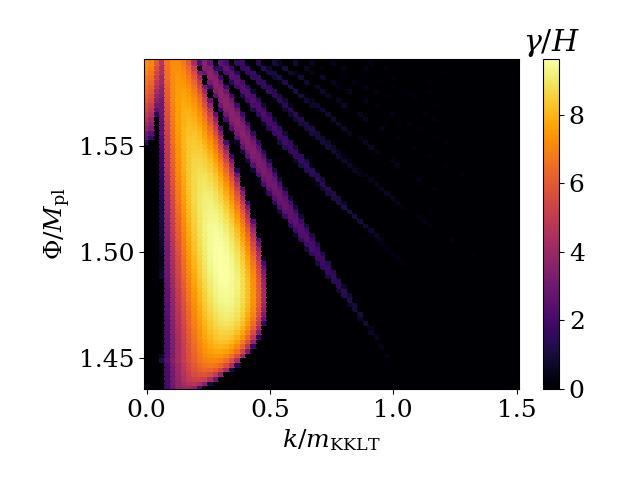}
    \includegraphics[width=0.49\linewidth]{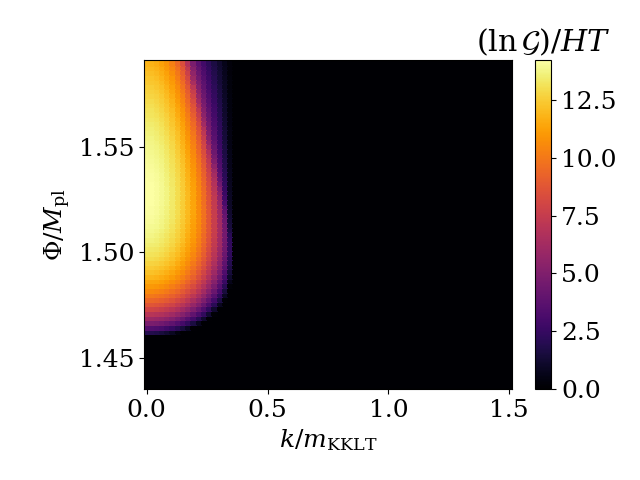}
    \caption{\textit{Left:} Floquet chart for self-resonance within the KKLT potential displayed in Fig.~\ref{fig:kklt-potn} as a function of modulus amplitude $\Phi/M_\text{pl}$ and wavenumber $k/m_\text{KKLT}$, where $m^2_{\rm KKLT}=V_{\rm KKLT}''(\varphi_0)$.
    \textit{Right:} Tachyonic growth rate approximated by integrating the tachyonic region as in Eq.~\ref{eq:logG_area}.}
    \label{fig:kklt-result}
\end{figure}

We display the stability maps coming from a full Floquet analysis for LVS in the left panel of Fig.~\ref{fig:LVS-result} and KKLT in the left panel of Fig.~\ref{fig:kklt-result}. 
The characteristic frequency is set by the mass of the modulus at the minimum of the potential, $m_\varphi^2\equiv V''(\varphi_0)$, which we note is several orders of magnitude smaller than $M_{\rm pl}$ in the potentials studied here.

Both bulk modulus potentials, LVS and KKLT, exhibit appreciable self-resonant instability bands.
Both Fig.~\ref{fig:LVS-result} and Fig.~\ref{fig:kklt-result} display broad regions with $\gamma/H\gtrsim \mathcal{O}(1)$ at low $k/m_\varphi$, indicating that self-resonant amplification of the modulus fluctuations can be efficient for a range of oscillation amplitudes. 
We also semi-analytically compute the contribution coming from tachyonic oscillations on the right panels of these figures. 
As expected from the examples in Sec.~\ref{setupgeneral}, the semi-analytical calculations more or less faithfully capture the growth rates near the tachyonic region away from $\Phi \sim 0$, while the full Floquet analysis also captures the contributions from parametric resonance near $\Phi \sim 0$.

These results are consistent with, and complement, earlier studies of moduli fragmentation and oscillon formation in string motivated potentials, notably the numerical analysis of~\cite{Antusch:2017flz}, who also find substantial self-resonant dynamics for KKLT moduli and for LVS volume moduli. 
The runaway, paired with the global minimum required for successful compactification, necessarily results in some region in between with $V''(\varphi)<0$, where tachyonic resonance may occur.
It is interesting to note that within the LVS framework, we find a growth parameter for self-interaction which is significantly larger than the growth parameter for a modulus coupled with axions \cite{WileyDeal:2025wgh,Leedom:2024qgr}.

Establishing the existence and location of resonance bands is a necessary first step, but it is not by itself sufficient to resolve cosmological questions such as the cosmological moduli problem; for that, one must also assess whether the ensuing nonlinear dynamics efficiently transfers energy into components that redshift faster than matter and/or efficiently populates light degrees of freedom.
These issues motivate the subsequent Section, where we investigate how light towers motivated by the Swampland can be incorporated as stochastic modulation of the Hill parameters governing the resonance structure.

\section{Swampland Distance Conjecture, Light KK Towers, and Stochastic Preheating}  \label{lvs-kk-swampland}

In this Section, we turn to a Swampland-inspired study of the second of the effective mass terms in Eq.~\ref{eq:intro_hill_schematic}: $\delta m_{\rm tower}^2(t)$.
Large-field potential structure, in addition to influencing the strength of the tachyonic region of oscillation, may not be trustworthy regions of field space, according to the SDC.
The EFT governing the potential breaks down and light scalars appear with exponentially small masses.
Here, we show that these light states can affect preheating dynamics despite the controlled field range of the oscillating modulus.

First we recapitulate the precise statement of the SDC for K\"ahler moduli following \cite{Corvilain:2018lgw}. For a low energy EFT defined at a point $Q$ in the moduli space, the SDC states that the EFT is only valid at finite distances around $Q$. Concretely, starting from $Q$ and moving to a different point $P$, an infinite tower of states becomes exponentially light, with the mass of the tower of states parametrically behaving as\footnote{We note that the parameter $\alpha$ defined below is totally separate from the parameter $\alpha$ used in the discussion of $\alpha$-attractor models.}
\begin{equation}\label{SDCdef}
    m(P) \,\, \sim \,\, m(Q) \, e^{-\alpha\, d(P,Q)}\,\,. 
\end{equation}
Here, $d(P,Q)$ is the geodesic distance between the points $P$ and $Q$. In our case, the point $Q$ will signify the location of the minimum of the potential in Fig.~\ref{fig:lvs-potn}, while the point $P$ will denote a generic location of the modulus during its oscillation.

The central working hypothesis of this Section  is that, after coarse-graining over one oscillation period, this tower of light modes can be modeled as an effective stochastic environment that modulates the coefficients of the Hill-type equations governing preheating. The validity of the stochastic regime, as opposed to a  full multi-particle treatment, is discussed in Appendix \ref{stoch-details-appendix}.

This idea reinterprets prior literature on noise and parametric resonance in reheating. The earliest analyses of reheating in the presence of noise considered a periodically driven mode equation with an additional random contribution to the effective mass \cite{Zanchin:1997gf, Zanchin:1998fj}. In particular, \cite{Zanchin:1997gf} showed that temporally uncorrelated homogeneous noise can increase the generalized Floquet exponent relative to the purely periodic case. They further extended this to spatially inhomogeneous noise, arguing  that the Floquet exponent is a non-decreasing function of the noise amplitude, and that the resulting growth can exceed the maximal noiseless Floquet exponent in appropriate regimes \cite{Zanchin:1998fj}. Stochasticity thus need not merely wash out resonance; depending on how it enters, it can also seed or enhance instability in regions that are stable in the deterministic problem.

\subsection{Light States and Stochastic Fluctuations in Hill's Equation}

Recent work by a subset of the current authors has revisited this problem in a form especially suited to our present purposes. In particular, ~\cite{Barrowes:2025cly} studied fluctuations in the parameters of Hill's equation directly, allowing the effective coefficients to vary from cycle to cycle and analyzing the resulting transfer matrix product with random matrix and stochastic methods. This framework makes transparent how modest fluctuations can broaden and shift instability bands, induce growth in otherwise stable regions, and, at larger variance, de-phase resonance or alter the mode structure of amplification. Importantly for the present paper, the authors explicitly motivate such fluctuations by couplings to additional light scalar fields, thereby providing a calculational bridge between multi-field dynamics and an effective stochastic Hill problem.

The general framework of fluctuating parameters in a Hill/Mathieu-type equation is developed as follows. Starting with a mode equation in Hill form, e.g.\ Mathieu’s equation $y''+[A+2qP(t)]y=0$, the cycle-indexed parameters $(A_i, q_i)$ are drawn from specified distributions to model the effect of many light fields. The solution of the $i$-th period is represented by a transfer matrix $M_i$ acting on the solution of the $(i-1)$th period, and the solution after $N$ periods is the product $\prod_{i=1}^N M_n$ acting on the initial state. The asymptotic growth rate is
\begin{equation}
\gamma T \;=\; \lim_{N\to\infty}\frac{1}{N}\,\log\Big\|\prod_{i=1}^N M_n\Big\|\,,
\label{eqn:noise-growth-rate}
\end{equation}
which can be analyzed using random matrix theory and stochastic methods. 
In the limit of zero noise, finding the growth rate amounts to finding the maximum eigenvalue of a single transfer matrix, and that is how we have generated all the noiseless Floquet maps.

Fluctuations can broaden and shift stability bands and, in some regimes, induce growth even where the noiseless system is classically stable. The net growth decomposes into a deterministic piece (a local average of surrounding growth rates on the map) and a noise-induced contribution (a non-negative “random walk” term in the logarithm of the matrix norm). As the variance of fluctuations increases, classic resonance regions are smeared and can either be enhanced or quenched depending on how the draws populate the unstable wedges.

Starting very generally, one can consider   a preheating condensate $\bar\varphi_2(t)$ (either the inflaton or a modulus) coupled to a tower of light fields $\chi_n$ through a potential $V(\varphi_2, \chi_n)$. The  masses $m_n$ of $\chi_n$ depend on a (possibly different) modulus $\varphi_1$ that parametrizes the canonically normalized displacement:
\begin{equation} \label{modKKEFT}
\mathcal{L}\supset -\frac{1}{2}m_n^2(\varphi_1)\,\chi_n^2
+ V_{\rm int}(\varphi_2, \chi_n)
+ V(\varphi_1) + V(\varphi_2) \,, 
\end{equation}
with
\begin{equation} \label{kkmasses}
m_n^2(\varphi_1)\in\Big\{m_{n0}^2\,e^{-2\alpha\,\Delta\varphi_1/M_{\rm pl}},\;\dots\Big\}.
\end{equation}
The potentials of the moduli $\varphi_1$ and $\varphi_2$ are $V(\varphi_1)$ and $V(\varphi_2)$, respectively. Here, $\alpha \sim \mathcal{O}(1)$ is a parameter that can range within certain model-dependent values (for example, \cite{Lust:2025auk}). 

\begin{figure}[htb!]
    \centering
    \includegraphics[width=0.49\linewidth]{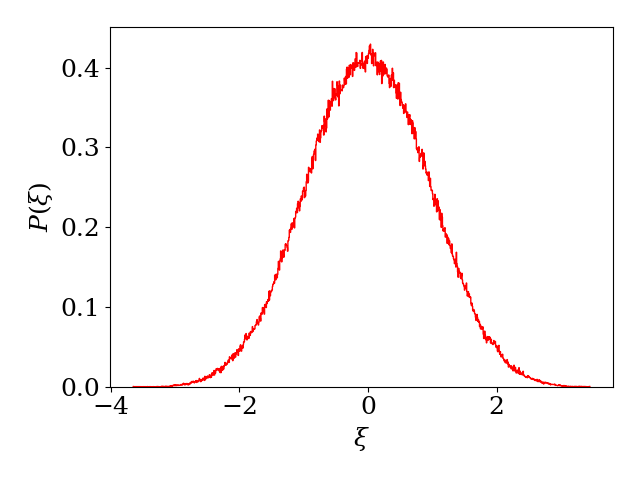}
    \includegraphics[width=0.49\linewidth]{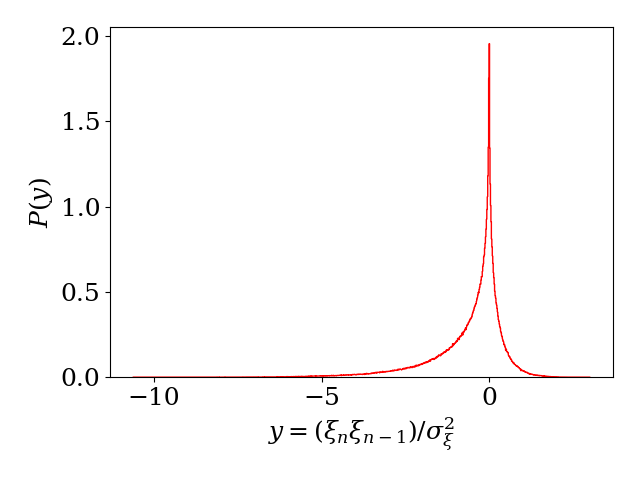}
    \caption{A tower of $20$ coherently oscillating scalars $\chi_i$ with masses $nm_{n0}e^{-\alpha\varphi_1/M_\mathrm{pl}}$, where integers $n\in(0,10^4)$ are sampled uniformly, and in this instance choosing $\alpha\varphi_1/M_\mathrm{pl}=10$, results in a normal distribution (left) of $\xi$ measured at each $\phi$-oscillation. The inter-period correlation (right) is centered close enough to zero that we can consider consecutive samples of $\xi$ to be uncorrelated. We use this Gaussian result, which we generally expect because of the Central Limit Theorem, to allow us to freely choose the variance of $\xi$ entering our numerical estimates. The Gaussian distribution of $\xi$ results in a chi-square distribution of $\xi^2$ with a single degree of freedom.}
    \label{fig:tower-dist}
\end{figure}

Even if one retains only a finite subset of modes below some conservative cutoff, the time dependence of $m_n(\bar\varphi_1(t))$ implies that the tower backreacts on the effective mass and couplings of the adiabatic fluctuation mode. After integrating out or coarse-graining over the rapidly responding tower sector over one oscillation, the fluctuation equation for the modulus perturbation $\delta\varphi_{2,k}$ naturally takes the form of a Hill equation with fluctuating parameters. In other words, as $\bar\varphi_1(t)$ oscillates or drifts, the ensemble $\{\chi_n\}$ induces time dependence in the coefficients of the adiabatic mode equation for $\delta\varphi_{2,k}$ through loop corrections and mean-field backreaction.

\subsection{Stochastic Preheating with LVS Example}

We work out the example of LVS in this Section, relegating a similar treatment for KKLT and blow-up moduli to Appendix \ref{stochKKLTblowup}.
We will mostly be interested in the  case when   $\varphi_1$ and $\varphi_2$ are identical, i.e., $\varphi_1 \equiv \varphi_2 \equiv \varphi$, and make a few brief comments about the non-identical case in Section~\ref{distinctmodulicase}. We have
\begin{equation} \label{modKKEFT2}
\mathcal{L}\supset -\frac{1}{2}m_n^2(\varphi)\,\chi_n^2
+ V(\varphi), \qquad 
m_n^2(\varphi)\in\Big\{m_{n0}^2\,e^{-2\alpha\,\Delta\varphi/M_{\rm pl}},\;\dots\Big\}.
\end{equation}
The $m^2_n(\varphi)\chi^2_n$ coupling in Eq.~\ref{modKKEFT2} becomes an interaction potential of the form 
\begin{align}
    V_{\rm int}(\varphi, \chi_n) = \frac{1}{2} m_{n0}^2e^{-2\alpha(\varphi-\varphi_0)/M_{\rm pl}}\chi_n^2
    \label{eqn:KK-mass}
\end{align}
We take LVS-type potentials for modulus  $V(\varphi)$. We will make use of the random walk term in the growth rate in the Floquet analysis of self-resonance in an LVS potential, with the deterministic part of the growth rate given in Fig.~\ref{fig:LVS-result}. The EOM of the $\varphi$ modes  gains an extra term
\begin{gather}
   \delta \ddot \varphi_{k} + \left(k^2+\partial_{\bar\varphi}^2V(\bar\varphi) +  \xi^2e^{-2\alpha(\bar\varphi-\varphi_0)/M_{\rm pl}}\right) \delta \varphi_{k} = 0 \quad \text{where} \quad
   \xi^2 \equiv 2\alpha^2 \sum_n\frac{m_{n0}^2\chi_n^2}{M_{\rm pl}^2}
    \label{eqn:stoch-eom}
\end{gather}
and $\xi=\xi(t)$ represents a continuous noise term.
Fig.~\ref{fig:tower-dist} demonstrates the distribution of the noise term $\xi$ for coherently oscillating $\chi_n$,
\begin{align}
    \chi_n(t)=X_n\cos(m_n t)\ ,
    \label{eqn:x-cos}
\end{align}
which turns out to be roughly Gaussian.
Equation~\ref{eqn:x-cos} is a reasonable heuristic picture of the behavior of scalar fields in the early Universe, which are generically displaced from their minima and coherently oscillate.
However, regardless of the exact choice of scalar field behavior, Gaussian noise is expected because of the Central Limit Theorem. 
Ref.~\cite{Barrowes:2025cly} finds that only six extra fields are required for the Gaussian limit to be reached; Fig.~\ref{fig:tower-dist} uses twenty, and many more are expected from the SDC.
Details concerning the validity of Gaussian noise are presented in Appendix~\ref{stoch-details-appendix}.

\begin{figure}
    \centering
    \includegraphics[width=0.49\linewidth]{map-H.png}
    \includegraphics[width=0.49\linewidth]{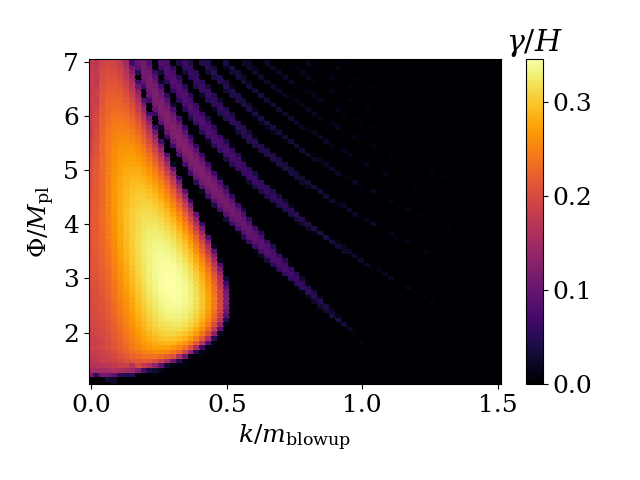}
    \includegraphics[width=0.49\linewidth]{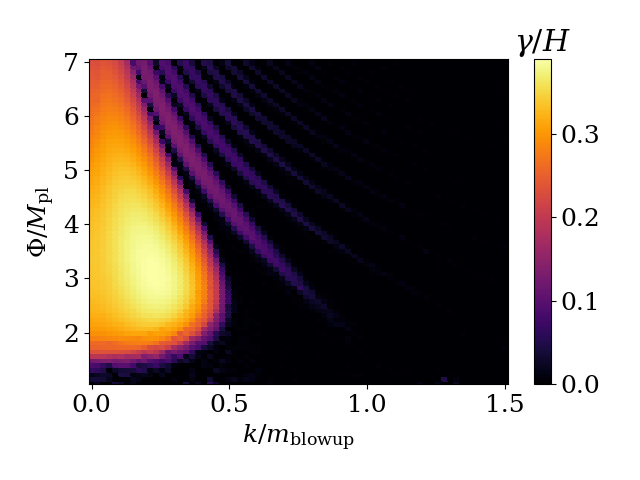}
    \includegraphics[width=0.49\linewidth]{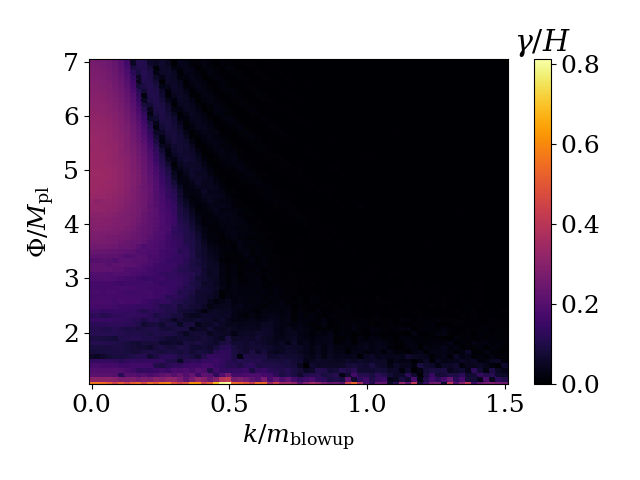}
    \caption{LVS self-resonance stability map with cycle-to-cycle stochasticity normally distributed with $\sqrt{\mathrm{Var}(\xi/m)}=0$ (top left), $0.5$ (top right), $1$ (bottom left), and $5$ (bottom right) appearing in the EOM in Eq. \ref{eqn:stoch-eom}. We use the modulus mass $m_{\rm LVS}^2=\partial_\varphi^2V_\mathrm{LVS}(0)$ as a characteristic scale. For moderate amounts of noise, stochasticity induced by the KK tower (with $\alpha=1$) washes away instability, but for a higher amount of noise we see a burst of instability for low $\Phi$. However, this noise-like approximation to backreaction may not be accurate enough for such a high $\xi$-variance.}
    \label{fig:lifted-stability}
\end{figure}

We utilize the correspondence between noise that is temporally continuous and noise that varies only between periods of $\varphi$, but is constant during each single period $i$~\cite{Barrowes:2025cly, Adams_2013}, allowing us to use Eq.~\ref{eqn:noise-growth-rate} to find the average growth rate per period.

\begin{figure}
    \centering
    \includegraphics[width=0.49\linewidth]{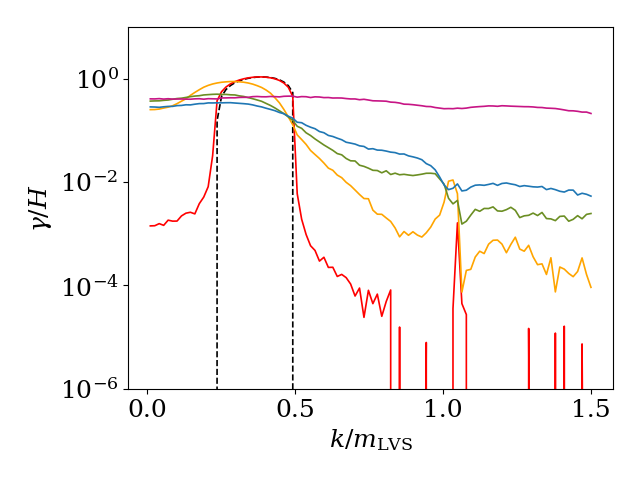}
    \includegraphics[width=0.49\linewidth]{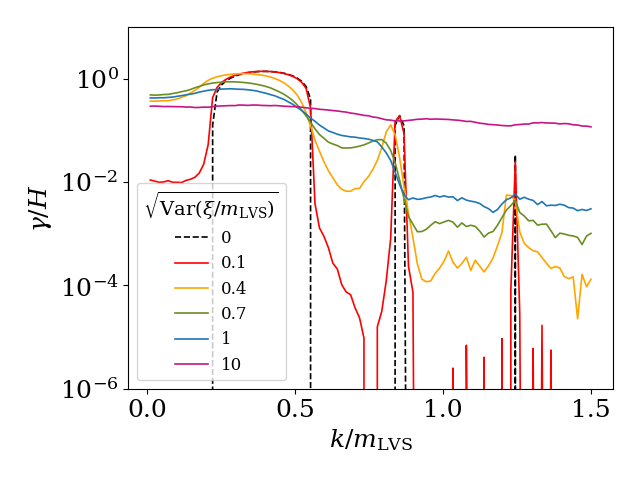}
    \caption{Slices of self-resonance stability in an LVS potential for $\Phi=0.2M_\mathrm{pl}$ (left) and $\Phi=0.4M_\mathrm{pl}$ (right), without fluctuations (dashed black) and a few choices of $\sqrt{\mathrm{Var} (\xi/m)}$ (solid colors) appearing in the EOM in Eq.~\ref{eqn:stoch-eom}. We use $\alpha=\frac{1}{2}$. We see a smearing of the resonance bands and a small lift from zero in previously stable regions. This lifting is possibly interesting, though likely not enough ($\gamma/H\lesssim\mathcal{O}(0.1)$ for realistic noise $\sqrt{\mathrm{Var} (\xi/m)}\leq1$) to result in significant particle production.}
    \label{fig:lvs-lines}
\end{figure}

Figs.~\ref{fig:lifted-stability}, \ref{fig:lvs-lines}, and \ref{fig:kk-preheating} summarize the main qualitative outcomes. For moderate stochasticity, the dominant effect is band smearing: sharp resonance wedges are broadened and their boundaries become diffuse, consistent with the expectation that cycle-to-cycle fluctuations de-phase coherent amplification. In addition, there is a small lift of the growth rate in some regions that are stable in the deterministic map, reflecting the fact that random draws occasionally place the system inside an unstable wedge even when the mean parameters lie outside. However, for the range of variances that we can motivate as ``reasonable'' in this effective description, the resulting enhancement of $\gamma/H$ is modest; in particular, while $\gamma/H$ can become positive in previously stable regions, it typically remains $\lesssim \mathcal{O}(0.1)$ for representative parameter choices. At very large variance the stochastic term can also produce bursts of IR instability, but such regimes should be interpreted cautiously. They likely correspond to dynamics in which the tower is no longer a mild perturbation but a strongly backreacting sector, and the simple noise approximation may not be applicable. We leave a more detailed exploration of this regime for the future.

\subsection{Energy Deposition into the KK Sector}

\begin{figure}
    \centering
    \includegraphics[width=0.49\linewidth]{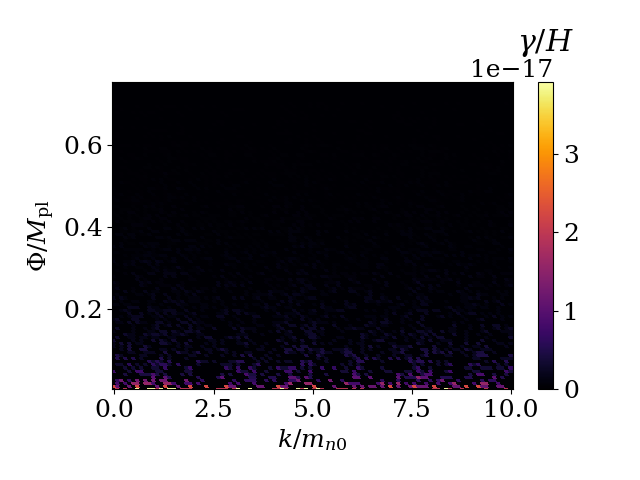}
    \includegraphics[width=0.49\linewidth]{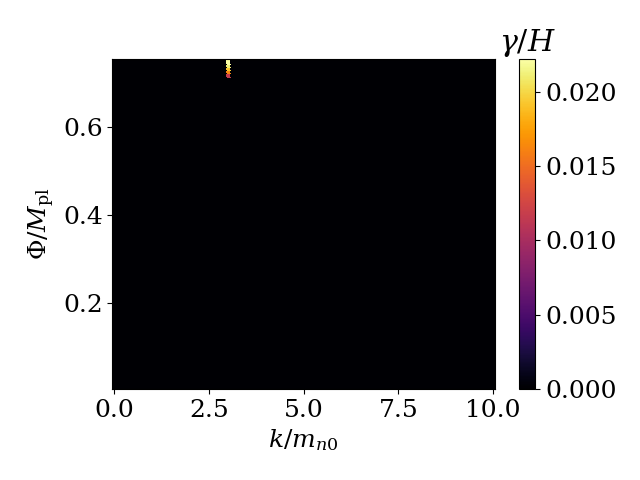}
    \caption{Floquet map of KK modes with $m_{n0}/M_\mathrm{pl}=1$ (left) and $10^{-3}$ (right), evolving with Eq.~\ref{eqn:tower-eom}, with $\varphi_1=\varphi_2=\varphi$, $\alpha=\frac{1}{2}\sqrt{\frac{3}{2}}$ (the maximum possible in the setting of \cite{Lust:2025auk}), and the zero mode $\bar\varphi$  oscillating in the LVS potential. The energy transfer is minimal. Any production of $\chi_n$ would increase the variance of the noise term in Eq.~\ref{eqn:stoch-eom}, improving the self-resonant effect of the modulus as shown in Fig.~\ref{fig:lifted-stability}. We do not see any Floquet instability of the KK modes except for a sliver with in the case of lower mass; this bit disappears for lower values of $\alpha$, so we can conclude that production of these KK modes would proceed perturbatively if at all.}
    \label{fig:kk-preheating}
\end{figure}

In the realm of fully multi-particle dynamics, one can ask whether the modulus $\varphi$ might deposit some of its energy into the KK tower. While we do not explore this in any detail in the current work, we can give some initial calculations. Each KK mode $n$ can be decomposed into Fourier components $k$, and each has the following EOM (neglecting cosmic expansion)
\begin{align}
    \ddot\chi_{n,k} + \left(k^2+m_{n0}^2e^{-2\alpha\bar\varphi/M_\mathrm{pl}}\right)\chi_{n,k} = 0\ ,
    \label{eqn:tower-eom}
\end{align}
where the oscillating zero mode $\bar\varphi$ in the LVS potential drives parametric resonance.

The stability map for this EOM for two choices of $k$ is shown in Fig.~\ref{fig:kk-preheating}.
In principle there can be some amplification of $\xi^2$ that appears in Eq.~\ref{eqn:stoch-eom}, improving the self-resonance efficiency of $\varphi$. A full nonlinear investigation of the backreaction of all fields on each other requires lattice simulations, which we reserve for the future.
However, it is clear from Fig.~\ref{fig:kk-preheating} that the minimal field content of a modulus and a tower of KK states is not enough to efficiently preheat the Universe.

\subsection{Two Distinct Moduli $\varphi_1$ and $\varphi_2$}\label{distinctmodulicase}

We finally consider the case where the moduli $\varphi_1$ and  $\varphi_2$ are different.
In this case, $\varphi_1$ is slowly rolling in its large field regime and controlling the KK tower mass while $\bar\varphi_2$ is oscillating about its minimum, attempting to reheat into $\delta\varphi_{2,k}$.
This scenario has no coupling between $\varphi_1$ and $\varphi_2$, so the tower of light fields has no effect on the self-resonant process of $\varphi_2$.

The EOM of the the actual KK modes $\chi_n$ is
\begin{align}
    \ddot\chi_{n,k} + \left(k^2+m_{n0}^2e^{-2\alpha\varphi_1/M_\mathrm{pl}}\right)\chi_{n,k} = 0\ .
    \label{eqn:chi-eom-12}
\end{align}
Here, the source term gets exponentially suppressed since $\varphi_1/M_{\rm pl}\gg1$, and its lack of oscillation removes any hope of resonant behavior.
Eq.~\ref{eqn:chi-eom-12} can be solved for a slowly-rolling modulus in an expanding Universe~\cite{Lust:2025auk}, where it is found that $\chi_n$ has a nonzero but non-growing solution, and the corresponding solution for $\varphi_1$ is explored as well.

\section{Conclusions}

We have explored the interplay between global asymptotics of the modulus potential and the efficiency of tachyonic preheating in Section~\ref{setupgeneral}. 
We find that the tachyonic branch of the refined de Sitter Conjecture supports the generic expectation of a period of tachyonic  self-resonance of moduli because of its lower bound on $\eta_V$ and a requirement that $V''$ be negative in some region of field space. However, while this diagnostic is necessary, it is not sufficient. The integrated gain also depends on the global shape of the potential, which decides the trajectory of the zero mode through the tachyonic region. We studied the interplay of local data and global asymptotics in several classes of potentials that occur in string compactifications: blow-up moduli, $\alpha$-attractor models, and bulk moduli with runaways. Future work will determine whether this diagnostic can serve as a useful criterion for identifying compactifications in which preheating efficiently reduces the abundance of moduli.

In Section~\ref{lvs-kk-swampland}, we explored the effect of light species on the preheating process, which appear because of the deviation of $\varphi$ from its minimum $\varphi_0$. 
For moderate stochasticity, sharp resonance wedges are broadened and their boundaries become diffuse, consistent with the expectation that  fluctuations from cycle to cycle de-phase coherent amplification. 
However, there is a small lift of the growth rate in regions that are stable in the deterministic map.
This aspect requires further investigation in the future through multifield lattice simulations. We also note that energy deposition into the KK tower is typically not very efficient.

\section{Acknowledgements}

We would like to thank Fred Adams, Tom Giblin, Aditi Shahani, and Scott Watson for enlightening discussion that improved the quality of this work. K.S. would like to thank the Aspen Center for Physics, which is supported by National Science Foundation grant PHY-2210452, for hospitality during the course of this work. The research activities of K.S. are supported in part by the U.S. National Science Foundation under Award No. PHY-2412671.
L.P. would like to thank the Leinweber Foundation for support at the University of Michigan, and the Center for Theoretical Physics at MIT for their hospitality.

\appendix

\section{LVS Potential} \label{lvsapp}

In this Section, we recapitulate the main ingredients that go into constructing the modulus potential in LVS. For more details, we refer to \cite{Cicoli:2023opf, Leedom:2024qgr, Antusch:2017flz}. We will consider compactifications with one bulk modulus  $\tau_b \equiv {\rm Re}\, T_b$ and one blow up cycle with modulus $\tau_s \equiv {\rm Re}\, T_s$. The modulus we will be most interested in here is $\tau_b$. We will assume that complex structure moduli are fixed by fluxes. The volume of the Calabi-Yau is given by
\begin{equation}
    \mathcal{V}
    \simeq
    \tau_b^{3/2}
    -
    \lambda
    \tau_s^{3/2}
\end{equation}
and, including leading-order $\alpha'$ corrections to the K\"{a}hler potential, we have 
\begin{align}
    \label{eq:lvsSuperPotential}
    W
    &
    =
    W_0
    +
    A_s
    \exp(
        -
        a_s
        T_s
    )
    +
    A_b
    \exp(
        -
        a_b
        T_b
    )
    \\
    \label{eq:lvsKahlerPotential}
    K
    &
    =
    -
    2
    \log
    \left[
        \mathcal{V}
        +
        \frac{
            \xi
        }{
            2
        }
    \right]
\end{align}
where 
\begin{equation}
    \xi
    =
    -
    \frac{
        \chi
        \,
        \zeta(3)
    }{
        2
        g_s^{3/2}
        (2\pi)^3
    }
    \sim 
    \mathcal{O}(1)
    .
\end{equation}
Here, $\xi$ depends on the Euler number $\chi$ of the Calabi-Yau manifold, and $g_s$, the string coupling, follows from the stabilization of the dilaton (i.e. $\xi \equiv \hat{\xi} ( S + \overline{S} )^{3/2}$ ). The scalar potential is obtained as
\begin{equation}
    \label{eq:lvsScalarPotential}
    V_F
    \simeq
    e^{K'}
    \left[
        \frac{8}{ 3 }
        \frac{
            a_s^2
            A_s^2
        }{
            \lambda
        }
        \sqrt{\tau_s}
        \frac{
            e^{-2 a_s \tau_s}
        }{
            \mathcal{V}
        }
        -
        4
        a_s
        A_s
        W_0 
        \tau_s
        \frac{
            e^{-a_s \tau_s}
        }{
            \mathcal{V}^2
        }
        +
        \frac{
            3
            \xi
            W_0^2
        }{
            4
            \mathcal{V}^{3}
        }
    \right]
\end{equation}
where we retain only terms up to $\mathcal{O}(\mathcal{V}^{-3})$ and $K'$ is the remaining K\"{a}hler potential which is assumed to be independent of the K\"{a}hler moduli, and we have stabilized the axion $c_s$ in the process.
Once the blow-up modulus $\tau_s$ is integrated out (i.e. stabilized), the potential becomes 
\begin{equation}
    V_F
    \simeq
    e^{K'}
    \left[
        -
        \frac{
            3
            \lambda
        }{
            2
            a_s^{3/2}
        }
        \log^{3/2}
        \left[
            \frac{
                4
                a_s
                A_s
                \mathcal{V}
            }{
                3
                W_0 
                \lambda
            }
        \right]    
        +
        \frac{
            3
            \xi
        }{
            4
        }
    \right]
    \frac{
        W_0^2
    }{
        \mathcal{V}^{3}
    }
    .
\end{equation}
However, this potential still has an AdS minimum and thus must be uplifted.
Here, we assume uplifting of the form 
\begin{equation}
    V
    \supset
    \frac{D}{\mathcal{V}^\gamma}
\end{equation}
where $D>0$ and $1\leq \gamma \leq 3$.
The potential is uplifted to a Minkowski minimum with  $V(\mathcal{V}=\langle \mathcal{V} \rangle) = 0$. The VEV and the uplifting coefficient $D$ can then be found from the  conditions $\frac{\partial V(\mathcal{V}=\langle \mathcal{V} \rangle)}{\partial \mathcal{V}} =0$ and $   V(\mathcal{V} = \langle \mathcal{V} \rangle) = 0$.

It is standard to rewrite this in terms of the canonical moduli through the field redefinition
\begin{equation}
    \tau_b = \exp(\sqrt{2/3} \phi).
\end{equation}
Expanding the potential around the (canonical) modulus VEV, $\phi = \langle \phi \rangle + \varphi$, produces the final potential used in the main body of the paper
\begin{equation}
    V
    \simeq
    \mathcal{A}
    \left[
        -
        \left(
            \sqrt{3/2} 
            \varphi
            +
            \mathcal{B}
        \right)^{3/2}
        +
        \mathcal{C}
    \right]
    e^{-3\sqrt{3/2} \varphi}
    +
    \mathcal{D}
    e^{-\sqrt{3/2} \gamma \varphi}
\end{equation}
where we have defined 
\begin{align}
    \mathcal{A}
    &
    \equiv
    W_0^2
    \frac{
        3
        \lambda
    }{
        2
        a_s^{3/2}
    }
    e^{-3\sqrt{3/2} \langle \phi \rangle}   
    \\
    \mathcal{B}
    &
    \equiv
    \sqrt{3/2} 
    \langle \phi \rangle
    +
    \log
    \left[
        \frac{
            4
            a_s
            A_s
        }{
            3
            W_0 
            \lambda
        }
    \right]
    \\
    \mathcal{C}
    &
    =
    \frac{
        2
        a_s^{3/2}
    }{
        3
        \lambda
    }
    \frac{
        3
        \xi
    }{
        4
    }
    \\
    \mathcal{D}
    &
    \equiv
    D
    e^{-\sqrt{3/2} \gamma \langle \phi \rangle}.
\end{align}
Note that $
    \langle \phi \rangle 
    =
    \sqrt{3/2}
    \log \langle \tau_b \rangle
    =
    \sqrt{2/3}
    \log \langle \mathcal{V} \rangle
$ and  $e^{K'} \simeq 1$.

\section{KKLT Potential}\label{KKLTblowup}

In this Appendix we show additional results of introducing stochasticity through KK towers for a KKLT and blowup modulus. We also introduce the parameters in the KKLT scenario used to define the modulus potential plotted in Fig.~\ref{fig:kklt-potn}. We largely follow the notation of \cite{Antusch:2017flz}, to which we refer for further details. 

The setup contains a single overall K\"ahler modulus \begin{equation} T=\tau+i\theta , \end{equation} with the complex structure moduli and the dilaton assumed to have already been integrated out at their supersymmetric minima. The resulting four dimensional effective supergravity is specified by 
\begin{equation} \frac{K}{M_{\rm pl}^2} = -3\log(T+\bar T)-K_{\rm cs}, \qquad \frac{W}{M_{\rm pl}^3} = W_0+A e^{-aT}. \label{eq:KKLT_KW_app} \end{equation} 
Here \(K_{\rm cs}\) denotes the vacuum value of the complex structure/dilaton contribution to the K\"ahler potential, while \(W_0\) is the flux superpotential after complex structure and dilaton stabilization. The non-perturbative term \(A e^{-aT}\) may arise, for example, from gaugino condensation on D7-branes or from Euclidean D3-brane instantons. In the gaugino condensation case one often has \(a=2\pi/N\), with \(N\) the rank of the condensing gauge group. The supersymmetric KKLT AdS minimum is determined by 
\begin{equation} 
D_T W=0. 
\end{equation} 
To obtain a metastable positive energy minimum, one adds an uplifting contribution, which for the purposes of ~\cite{Antusch:2017flz} is taken to scale schematically as 
\begin{equation} V_{\rm up}\propto \left(\frac{T+\bar T}{2}\right)^{-2}. \label{eq:KKLT_uplift_app} \end{equation} 
Different microscopic uplift mechanisms may produce different powers of \(T+\bar T\); the above choice is a representative convention. The F-term part of the scalar potential for the real modulus \(\tau=\mathrm{Re}\,T\), after fixing the axion at its minimum, is written as 
\begin{equation} \frac{V_{\rm F}}{M_{\rm pl}^4} = \frac{e^{K_{\rm cs}}}{6\tau^2} \left[ a A^2 (3+a\tau)e^{-2a\tau} -3a A W_0 e^{-a\tau} \right], \label{eq:KKLT_Fterm_app} \end{equation} 
with the uplift term added separately when constructing the metastable dS minimum. In the main text, when referring to the KKLT potential, we mean the resulting uplifted single field potential along the real direction after the axion and all other moduli have been fixed. 

The canonically normalized real field \(\phi\) associated with the volume modulus is 
\begin{equation} \frac{\phi}{M_{\rm pl}} = \frac{\sqrt{3}}{2}\log(T+\bar T) = \frac{\sqrt{3}}{2}\log(2\tau). \label{eq:KKLT_canon_app} \end{equation} 
Thus a potential written as \(V(\tau)\) can be converted to a potential for the canonical field by 
\begin{equation} \tau(\phi)=\frac{1}{2}\exp\!\left(\frac{2\phi}{\sqrt{3}M_{\rm pl}}\right). \end{equation} 
 For the numerical studies of~\cite{Antusch:2017flz}, the representative parameter range was chosen as 
 \begin{equation} 10^{-12}\leq W_0\leq 10^{-5}, \qquad 1\leq A\leq 10, \qquad 1\leq a\leq 2\pi, \label{eq:KKLT_parameter_range_app} \end{equation} 
 with \(e^{K_{\rm cs}}\) set to unity. The subsequent lattice simulations show that the dominant growth mechanism in the KKLT examples is parametric self-resonance rather than the initial tachyonic passage alone, although the tachyonic region and barrier structure are important for organizing the allowed initial displacements.

\section{Stochasticity for KKLT and Blowup Moduli}\label{stochKKLTblowup}

This Appendix displays results of adding stochasticity to the KKLT and blowup systems, where the KK tower masses now determined by these moduli.

\begin{figure}
    \centering
    \includegraphics[width=0.49\linewidth]{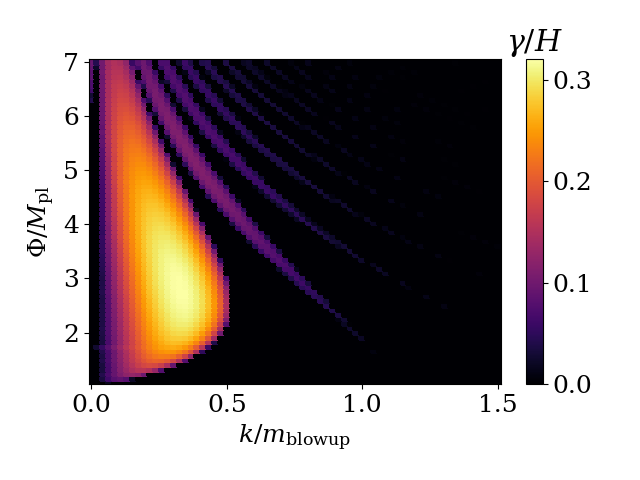}
    \includegraphics[width=0.49\linewidth]{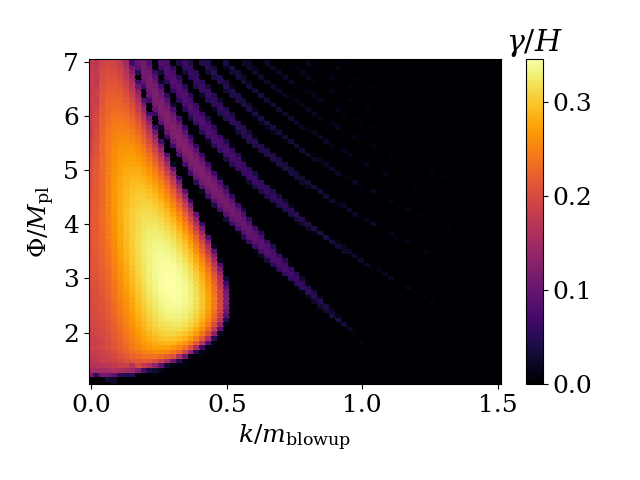}
    \includegraphics[width=0.49\linewidth]{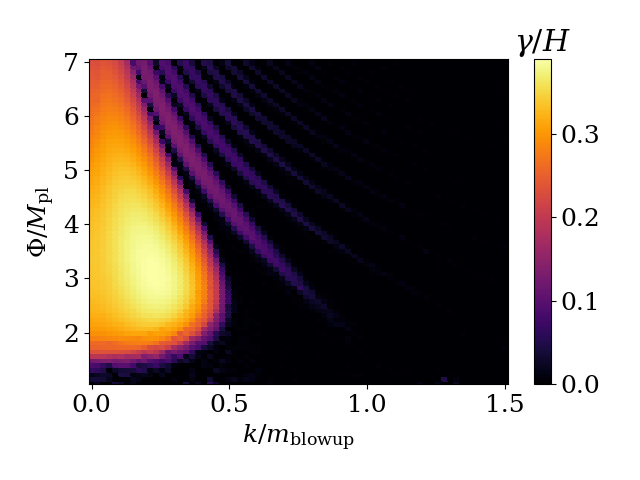}
    \includegraphics[width=0.49\linewidth]{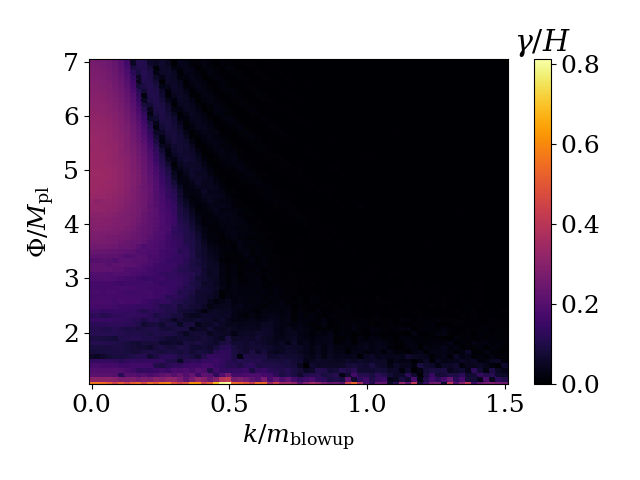}
    \caption{Blowup modulus self-resonance stability map with cycle-to-cycle stochasticity normally distributed with $\sqrt{\mathrm{Var}(\xi/m)}=0$ (top left) $0.5$ (top right), $1$ (bottom left), and $5$ (bottom right) appearing in the EOM in Eq. \ref{eqn:stoch-eom}. We use the modulus mass $m_{\rm blowup}^2=\partial_\varphi^2V_\mathrm{blowup}(\varphi_0)$ as a characteristic scale. For moderate amounts of noise, stochasticity induced by the KK tower (with $\alpha=1$) washes away parametric instability while slightly enhancing some regions of tachyonic instability. For higher noise, a burst of particle production starts to appear for $\Phi$ close to the VEV, although this size of noise is large and should be studied with higher fidelity. In all cases, lower energy modes $k\lesssim0.1$ gain more access to resonant decay as a result of noise.}
    \label{fig:blowup-stoch}
\end{figure}

\begin{figure}
    \centering
    \includegraphics[width=0.49\linewidth]{new-map-H.png}
    \includegraphics[width=0.49\linewidth]{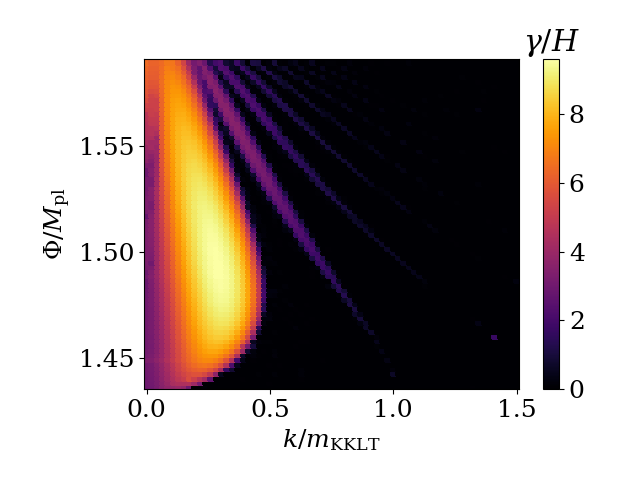}
    \includegraphics[width=0.49\linewidth]{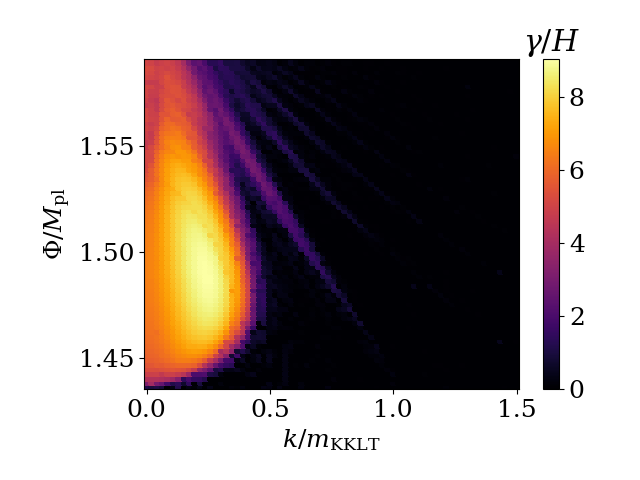}
    \includegraphics[width=0.49\linewidth]{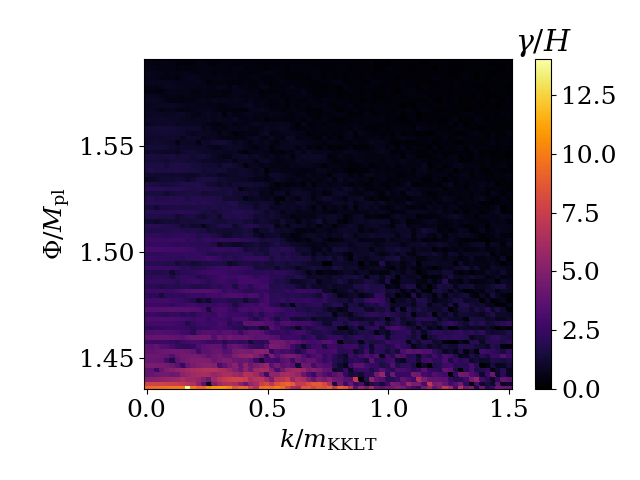}
    \caption{KKLT modulus self-resonance stability map with cycle-to-cycle stochasticity normally distributed with $\sqrt{\mathrm{Var}(\xi/m)}=0$ (top left) $0.5$ (top right), $1$ (bottom left), and $5$ (bottom right) appearing in the EOM in Eq. \ref{eqn:stoch-eom}. We use the modulus mass $m_{\rm KKLT}^2=\partial_\varphi^2V_\mathrm{KKLT}(\varphi_0)$ as a characteristic scale. For moderate amounts of noise, stochasticity induced by the KK tower (with $\alpha=1$) acts mainly to induce more resonance for $k/m_{\rm KKLT}\lesssim0.1$, but for a higher, possibly unrealistic amount of noise we see the resonance structure give way to enhanced perturbative decay.}
    \label{fig:kklt-stoch}
\end{figure}

As in the LVS case, the tower and modulus are coupled by the tower masses Eq.~\ref{eqn:KK-mass}, and the EOM of the modulus Fourier modes is Eq.~\ref{eqn:stoch-eom}, but now with different choices of modulus potential $V(\varphi)$.
The blowup modulus potential is given by Eq.~\ref{eq:KMII_potential} and shown in Fig.~\ref{fig:blowup-V}, and the self-resonant and stochastic results are shown in Fig.~\ref{fig:blowup-stoch}. The KKLT potential is shown in Fig.~\ref{fig:kklt-potn}, and the self-resonant and stochastic results are shown in Fig.~\ref{fig:kklt-stoch}.

We see similarities to the addition of noise in the LVS system: particle production is enhanced for low wavenumber and fades at the edge of the broad resonance band.

 \section{Details of Stochasticity in the Hill's Equations}  \label{stoch-details-appendix}

We make a few comments on the regime in which it is valid to model the tower of light modes as an effective stochastic environment modulating the coefficients of the  equations governing preheating. The stochastic treatment is expected to work when a large number of  underdamped KK modes below the EFT cutoff contribute comparably to the average backreaction per cycle. Conversely, the stochastic approximation breaks down when the tower is dominated by a small number of coherent modes. In other words, if only a few modes dominate the tower of KK states and their oscillation frequencies are commensurate with the oscillation frequency of the modulus $\varphi$, then the correct approximation to use is a quasi-periodic multifield treatment rather than a stochastic one. While such a treatment is interesting, it is beyond the scope of the current work.

One can make these statements precise in the context of Eq.~\ref{eqn:stoch-eom}. We begin with treating the term
\[
\xi^2 e^{-2\alpha(\bar\varphi-\varphi_0)/M_{\rm pl}},
\qquad
\xi^2\equiv
\frac{2\alpha^2}{M_{\rm pl}^2}
\sum_n m_{n0}^2\chi_n^2 ,
\]
as introduced in Eq.~\ref{eqn:stoch-eom} as a deterministic contribution in the full multifield system. To model this deterministic contribution as a stochastic term, one needs to consider  $\xi^2$ and take the time average over one oscillation cycle of the modulus, say the $i$th cycle (the corresponding quantity is therefore $\xi^2_i$):
\begin{equation}
\xi_i^2
\equiv
\frac{1}{T_\phi}
\int_{iT_\phi}^{(i+1)T_\phi}dt\,
\frac{2\alpha^2}{M_{\rm pl}^2}
\sum_{n=1}^{N}m_{n0}^2\chi_n^2(t),
\qquad
i=\text{cycle index},
\label{eq:xi_cycle_average}
\end{equation}
Here $T_\varphi$ is the oscillation period of the modulus $\varphi$ and $N$ denotes the number of KK modes. The next step is to write $\xi^2_i$ as an average contribution $\langle \xi^2\rangle$ over the contributions of the KK modes $n=1\ldots N$, plus a fluctuation $\delta\xi_i^2$ about the mean 
\begin{equation}
\xi_i^2=\langle \xi^2\rangle+\delta\xi_i^2 .
\label{eq:xi_mean_fluctuation}
\end{equation}
The mean contribution \(\langle\xi^2\rangle\) gives a deterministic shift of the effective mass,
while \(\delta\xi_i^2\) is treated as the stochastic input. Thus the equation for
\(\delta\varphi_k\) becomes
\begin{equation}
\delta\ddot\varphi_k+
\left[
k^2+V''(\bar\varphi)
+
\langle\xi^2\rangle e^{-2\alpha(\bar\varphi-\varphi_0)/M_{\rm pl}}
+
\delta\xi_i^2 e^{-2\alpha(\bar\varphi-\varphi_0)/M_{\rm pl}}
\right]\delta\varphi_k=0,
\quad
t\in[iT_\varphi,(i+1)T_\varphi).
\label{eq:xi_stochastic_hill}
\end{equation}
which we are writing in the more compact form given in Eq.~\ref{eqn:stoch-eom}.

The stochastic approximation is justified only in a restricted regime. First, the active \(\chi_n\) modes
must be genuine dynamical degrees of freedom of the EFT:
\begin{equation}
H\ll \omega_{\chi,n}(t)\ll \Lambda_{\rm EFT},
\qquad
\omega_{\chi,n}^2(t)
\equiv
\frac{k_\chi^2}{a^2}
+
m_{n0}^2 e^{-2\alpha(\bar\phi-\phi_0)/M_{\rm pl}} .
\label{eq:chi_active_condition}
\end{equation}
The lower bound ensures that the \(\chi_n\) are underdamped rather than frozen by Hubble
friction, while the upper bound keeps them inside the EFT. Modes with
\(\omega_{\chi,n}\lesssim H\) behave as slowly varying backgrounds, not as stochastic
fluctuations.

Secondly, for any given cycle $i$,  the Gaussian approximation requires that many weakly weighted contributions appear in $\delta\xi^2_i$, with no single KK mode dominating. We will denote $\delta\xi^2_i \equiv \delta\xi^2$ below, with the understanding that we are dealing with the $i$th cycle.  Formally, one can decompose 
\begin{equation}
\delta\xi^2=\sum_{n=1}^{N} \delta\xi_{n}^2 ,
\label{eq:xi_sum_modes}
\end{equation}
and define
\begin{equation}
N_{\rm eff}
\equiv
\frac{\left(\sum_n \sigma_n^2\right)^2}{\sum_n\sigma_n^4},
\qquad
\sigma_n^2\equiv
{\rm Var}(\delta\xi_{n}^2).
\label{eq:xi_neff}
\end{equation}
The central limit approximation then  requires
\begin{equation}
N_{\rm eff}\gg 1,
\qquad
\frac{\max_n\sigma_n^2}{\sum_m\sigma_m^2}\ll 1 \,\,,
\label{eq:xi_lindeberg}
\end{equation}
with the second condition ensuring that no single mode dominates. 

Finally, we demand that the tower of KK states constitutes an environment rather than a dominant component of the dynamics. The energy density in KK modes should be subdominant: $
\rho_\chi\ll\rho_\phi$. Moreover, defining
\begin{equation}
\epsilon_\xi
\equiv
\frac{\sigma_{\xi^2}}{\Delta m_{\rm det}^2},
\qquad
\sigma_{\xi^2}^2\equiv {\rm Var}(\delta\xi^2),
\qquad
\Delta m_{\rm det}^2
\equiv
\max_{\rm cycle}V''(\bar\varphi)-
\min_{\rm cycle}V''(\bar\varphi),
\label{eq:xi_noise_strength}
\end{equation}
the controlled stochastic regime is
\begin{equation}
\epsilon_\xi\lesssim 1
\label{eq:xi_backreaction_condition}
\end{equation}
 For \(\epsilon_\xi\gg1\), or when \(\rho_\chi\) becomes
comparable to \(\rho_\phi\), the stochastic approximation is no longer controlled and one
should instead solve the coupled multifield dynamics, ultimately with lattice simulations. We leave this direction for the future.

\bibliography{ref}

\end{document}